\documentclass[12pt, draftclsnofoot, onecolumn]{IEEEtran}

\usepackage[switch]{lineno}

\usepackage{amsmath}
\usepackage{graphicx}
\usepackage{subfigure}
\usepackage{multicol} 
\usepackage{epstopdf} 
\usepackage{epsfig} 
\usepackage{multirow}
\usepackage{bm}
\usepackage{array}
\usepackage{color}
\usepackage{float}
\usepackage{amsfonts}
\usepackage{caption}

\ifCLASSINFOpdf
\else
\fi
\begin{document}
%
\title{Coding Capacity of Purkinje Cells with Different Schemes of Morphological Reduction }

\author{\IEEEauthorblockN{Lingling~An\dag, Yuanhong~Tang\dag,  Quan~Wang\dag, Qingqi~Pei\dag, Ran~Wei\dag, Huiyuan~Duan\dag, Jian~K.~Liu\ddag \\}
	\IEEEauthorblockA{\dag  School of Computer Science and Technology, Xidian University, Xi'an, China\\
		\ddag Centre for Systems Neuroscience,  
        Department of Neuroscience, Psychology and Behaviour, University of Leicester, Leicester, UK\\
		Correspondence:: an.lingling@gmail.com, jian.liu@leicester.ac.uk
        }
}


\maketitle

\begin{abstract}

The brain as a neuronal system has very complex structure with large diversity of neuronal types. The most basic complexity is seen from the structure of neuronal morphology, which usually has a complex tree-like
structure with dendritic spines distributed in branches. For simulating a large-scale network with spiking neurons, the simple point neuron, such as integrate-and-fire neuron, is often used. However, recent experimental evidence suggests that the computational ability of a single neuron is largely enhanced
by its morphological structure, in particular, by various types of dendritic dynamics. As morphology reduction of detailed biophysical models is one of classic questions for systems neuroscience, much effort has been taken to simulate a neuron with a few compartments to include the interaction between soma and dendritic spines. 
Yet, novel reduction methods are still needed to deal with complex dendritic tree.
Here by using ten individual Purkinje cells of the cerebellum from three species of guinea-pig, mouse and rat,
we consider four types of reduction methods and study their effects on the coding capacity of Purkinje cells in terms of firing rate, timing coding, spiking pattern, and modulated firing under different stimulation protocols. We find that there is a variation of reduction performance depending on individual cells and species, however, all reduction methods can preserve, to some degree, firing activity of the full model of Purkinje cell. Therefore, when stimulating large-scale network of neurons, one has to choose a proper type of reduced neuronal model depending on the questions addressed.
Among these reduction schemes, Branch method, that preserves the geometrical volume of neurons, can achieve the best balance among different performance measures of accuracy, simplification, and computational efficiency, and reproduce various phenomena shown in the full morphology model of Purkinje cells. Taken together, these results suggest that Branch reduction scheme seems to provide a general guideline for reducing complex morphology into a few compartments without loss of basic characteristics of firing property of neurons.


\end{abstract}

%
\IEEEpeerreviewmaketitle

\section{Introduction}
A single neuron is thought as the basic computation unit in the complex neuronal system. Understanding the computations and dynamics of single neurons is an integral part of explaining their functional roles within a large neural network. However, there is a complex dendritic structure for a single neuron's morphology. The question is what are the key properties of neuronal morphology for understanding the complex process such as information processing and dynamic behavior emerged from a network of neurons. How one can reduce the whole morphology but still keeping these essential dynamics? To address these questions, there are a number of methods and models developed to describe firing properties of single neurons at different levels of abstractions and reductions.  
A simple, perhaps the most compact version of a neuronal model is the so-called integrate-and-fire (IAF) model~\cite{Lapicque1907}. The IAF model of neuronal dynamics is considered as a simple way to explain certain aspects of neuronal behaviors~\cite{Burkitt2006A,Burkitt2006A1}. In particular, it has been used mostly for simulation of large-scale neuronal networks. However, with the advancement of experimental techniques for characterizing fine dynamics of neurons, there is an increasing volume of evidence that the IAF approach cannot capture the dynamics and computations of real neurons~\cite{Mainen1996Influence,  Koch1999Biophysics, Poirazi2003Arithmetic, Ostojic2015Neuronal,amsalem2018efficient}, in which neuronal morphology, ion channels, and synapse distributions, all affect the activity of neurons~\cite{Herz2006Modeling}. The classical Rall cable theory is useful to understand the contribution of neuronal spatial structure to its function and dynamics~\cite{Rall1959Branching}. By extending the Rall cable theory, the real neuronal morphology can be reconstructed based on the anatomy to ensure that its electrophysiological characteristics are unchanged. Although the detailed model can describe the dynamics of individual neurons very well, their high dimensionality and complex spatial structure make the calculation expensive and unsuitable for large-scale network simulation.

The classical Hodgkin-Huxley single-compartment model ignores the spatial structure of neurons to mainly explain the ionic mechanisms and how action potentials are initiated and propagated in neurons~\cite{Hodgkin1990A}. 
However, the recent theoretical work shows that the single somatic point neuron is not enough to capture the detailed firing activities of Purkinje cell observed in experiments~\cite{Ostojic2015Neuronal}. The neuronal model with at least two compartments of soma and dendrite is necessary. 
Similarly, the reduced model with only one or a few dendritic compartments is usually sufficient to understand detailed neuronal activities~\cite{Brown2011Virtual, Armin2012Automated,amsalem2018efficient, Marasco2013Using}. For large-scale network studies, reduced models provide a good balance between biological activity and computational efficiency~\cite{Herz2006Modeling}. However, what kind of reduction schemes for simplifying the whole dendritic morphology is still less well understood~\cite{amsalem2018efficient, Marasco2013Using}.

Here we address this question by focusing on the Purkinje cells (PCs) of the cerebellum. The cerebellum, as one of the most well-studied brain areas, traditionally plays an essential role for motor control~\cite{Ito1984The,Zeeuw1997Volume, Amir2011Cerebellum, Manto2012Consensus} and learning of vestibular-ocular reflex~\cite{Du1995Learning, Hirata2012Direct, Blazquez2004The} and eyelid reflex regulation~\cite{Koekkoek2003Cerebellar, Jim2004Role}. In recent years, a large number of studies have shown that the cerebellum is also involved in the processing of information such as cognition, language, attention, and memory~\cite{Wolf2009Evaluating,Wagner2017Cerebellar,Ito2008Control,Strick2009Cerebellum,Tsai2012Autistic,Bostan2018The,Raymond2018Computational}.

PCs, as the only output neurons in the cerebellum, is an indispensable component in the mechanism of synaptic plasticity in cerebellar learning. PCs receive parallel fiber (PF) input that generates high-frequency simple spikes (SSs) to predict ongoing movements~\cite{Loewenstein2005Bistability}. The high-frequency SS discharge of PCs encodes the information about movements,  including performance errors and kinematics~\cite{Popa2016The, Robinson2001The}. Importantly, SS modulation both leads and lags behavior, which means that individual PCs may carry predictive and feedback signals about motor commands and corresponding behaviors~\cite{Chen2016The, Hewitt2015Changes, Popa2015Predictive, Streng2018Modulation}. 
PC has a tree-like morphology structure with an intricately elaborate dendritic arbor which can be characterized by a large number of dendritic spines distributed in its dendritic branches.  
Given such a sophisticated dendritic structure,  it is necessary to find a suitable level of abstraction with a reduced morphological structure to understand the working mechanisms of PC for integration of PF synaptic inputs. 

In a previous study~\cite{Marasco2013Using}, a reduction method based on Strahler analysis of neuron morphology was applied to PC with arbitrary dendritic distribution of ion channels and synaptic inputs and without any fitting or tuning procedures. In particular, this reduction method can ensure somatic membrane potential trace accurate while reducing the runtime of simulation significantly. In this study, we systematically model the morphology of PC from three species of guinea-pig, mouse and rat, and investigate the effect of four different reduction schemes on the coding capacity of PC, in terms of firing rate, spike timing, spike pattern, and modulation of firing amplitude and phase, with several types of stimulation protocols of PF synaptic inputs, including Poisson and modulated renewal process. 

We find that the PC firing rate coding, which is the input-output relationship of the PC firing activity under different frequencies, is slightly different in low frequency stimuli but significantly different in high frequency stimuli at a millimeter-scale reaction under different reduction schemes of morphology. 
The PC spike timing coding quantified by the inter spike intervals and regular/irregular spike patterns are well captured by different reduction schemes. In addition, for modulated stimulus with a frequency following a sinusoidal modulation, the phase and amplitude of the PC response curve are also well described by different reduction schemes. However, there is a variation of coding performance across different reduction schemes for PCs from different species. 
Among these reduction schemes, Branch method, which keeps the feature of the geometrical volume of a neuron, can achieve a good balance between different performance measures of accuracy, simplification, computational efficiency, and spike shape change of morphology reduction, and reproduce various phenomena shown in the full model with the whole PC morphology. Thus, these results suggest that Branch reduction scheme could serve as a general guideline for reducing complex morphology into a few compartments without loss of basic characteristics of firing property of neurons.

\section{Methods}

\begin{figure*}[tbpt]	
	\centering	
		\includegraphics[width=\columnwidth]{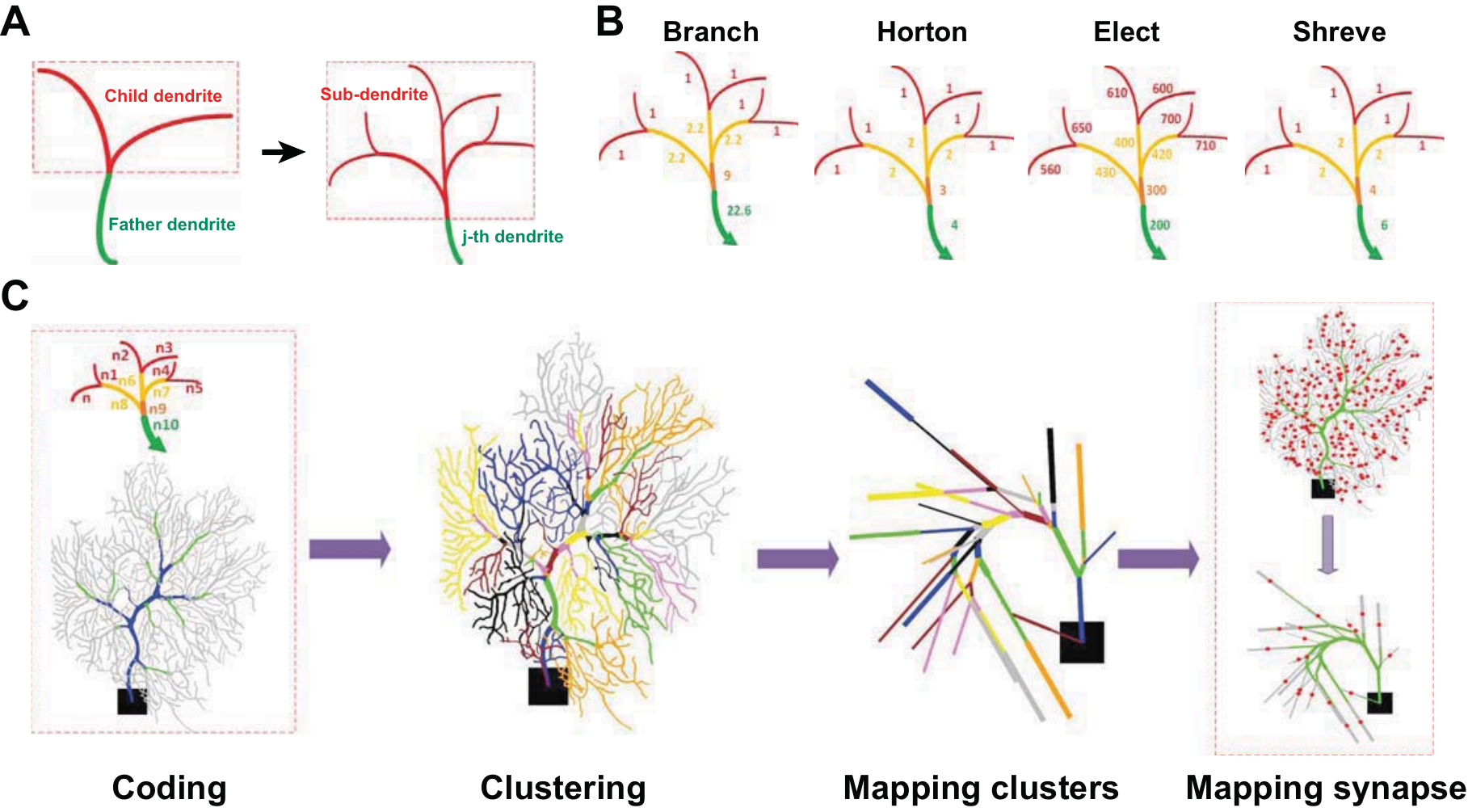}
	\caption{ Schematic view of reduction process. (A) Illustration of a dendritic field. Each sub-branch has a father dendrite with several child dendrites. Each of them can be indexed as the $j$-$th$ dendrite (green) with a set of sub-dendrites (red).
     (B) Illustration of different coding schemes of neuronal morphological structure. Levels of dendrites are colored differently with terminal dendrites as red and non-terminal dendrites as other colors. The number of each dendrite is the order value obtained by different coding schemes. For instance, for Branch model, the order value of non-terminal dendrites is the sum of all their sub-dendrite order values plus a weight as the number of sub-dendrites divided by 10. Thus, the order values of all terminal dendrites are 1. At the next level, the dendrites have a order value as 2.2 that is the sum of 2, from all their sub-dendrite order values, with 0.2 as a weight of the number of sub-dendrites divided by 10. So the final order value at this level is 2.2. This process moves from the terminal or non-terminal dendrites level by level to get all the order values. Note Elect method operates in an opposite way. 
       (C) Reduction process. Coding: The 1st step is to encode morphological structure and distinguish different functional areas (Blue, Trunk. Green, Smooth. Gray, Spiny) of a neuron. Clustering: the 2nd step is to set clusters according to the coding number. Mapping clusters: the 3rd step is to map each cluster into a single compartment. Mapping synapses: the 4th step is to map synaptic locations in the reduced model. The red dots indicate synaptic locations in the spiny dendrites.
        }
	\centering
	\label{cluster}
\end{figure*}

\subsection{Neuronal morphology reduction}

A reduction method is to map the full morphological structure into an equivalent reduced model with much less dendritic compartments. As a typical tree-like structure, one has to identify the father and child dendrites, and mark them with a set of graphic notations (see Fig.~\ref{cluster}(A)). For the markers, one needs to set up a coding scheme such that there is an order value for each selected section or area as illustrated in Fig.~\ref{cluster}(B). Here we propose four different coding schemes where the order values are determined differently. The motivations of these coding schemes are inspired by general analysis of river networks ~\cite{Horton1945EROSIONAL, Shreve1967Infinite}. 

The main feature of Shreve and Horton methods is to identify and classify river types based on the importance of rivers in water networks.
Specifically, Shreve coding is to define a river without tributaries as level 1, and other river levels are obtained by adding their tributary levels. The characteristics of the Shreve method are similar to the calculation of confluence, and have a relatively large reference value in some simulations of hydrological flow and sediment volume. Horton coding is to define a river without tributaries as level 1, and other river levels are obtained by maximum level of its tributary. Horton code mainly reflects the hierarchical relationship in the water network and the depth of the river subtree. 
The Purkinje cell dendritic structure is similar to the river network, and these two methods can be used to classify the dendrites and then merge and simplify the unimportant dendrites.
As a result, the different functional regions of PCs can be characterized by their order values.

Before setting the order value, the PC tree-like morphological structure can be divided into terminal and non-terminal dendrites. We use $E$  and  $O$ to represent the set of terminal dendrites and non-terminal dendrites, respectively. Specifically, $E =\{e_1, e_2, \cdots, e_i\}$ is terminal dendrites set,  $ e_i$ is the $i$-$th$ dendrite in the set $E$,   $N_{e,i}$ is the order value of  the $i$-$th$ dendrite in set $E$. 
$O=\{o_1,o_2,….. o_j\}$ is non-terminal  dendrites set,  $o_j$ is the $j$-$th$ dendrite in $O$, $ N_{o,j}$ is the order value of the $j$-$th$ dendrite in $O$. Then according to the order value $N$, $N=N_{e,i}$ or $N=N_{o,j}$, the dendrites can be classified as spiny, smooth, or trunk, as following: \begin{equation}
dend\in \left\{ \begin{array}{ll}
Spiny, & \textrm{if }\quad N \leq {s1} \\
Smooth, & \textrm{if}\quad  s1<N<s2\\
Trunk ,& \textrm{if}\quad N\geq{s2}
\end{array} \right.
\label{classify}
\end{equation}
where $s1$ and $s2$ are the best threshold values for dividing the functional areas of spiny, smooth, and trunk dendrites. We tested several different values of $s1$ and $s2$, and determined the best values that can achieve the balance between a higher reduction and a better overall accuracy. These values of chosen for each PC are given in Table~\ref{best_threshold}.

In this study, we used four different coding schemes as below.

\subsubsection{Branch scheme} In this scheme, the order values of dendrites are determined by the branches of each dendrite. The order values of terminal dendrites are fixed as 1, i.e., $N_{e,i}=1$, then the non-terminal dendrites order values are given as:
\begin{equation}
N_{o,j}=\sum_{k=1}^{n} N_{o,j_k}+weight
\end{equation}
$N=\{N_{o,j_1},N_{o,j_2},\cdots N_{o,j_n}\}$ is the order value of sub-dendrites of the $j$-$th$ dendrite in set  $O$, $j_k$ is $k$-$th$ sub-dendrite of the $j$-$th$ dendrite, $n$ is the number of the sub-dendrites of the $j$-$th$ dendrite, and $weight=n/10$, where the factor 10 is the best value for trade-off between simplification and accuracy of the modelling in our case. 

\subsubsection{Horton scheme} In this scheme, the dendritic tree is encoded by Horton analysis~\cite{Horton1945EROSIONAL}. We set the order value for each section. Terminal dendrites of set $E$  have the order value as 1, $ N_{e,i} =1 $. The order value for non-terminal dendrites can be set as:
\begin{equation}
N_{o, j}=max(N_{o, j_1}, N_{o, j_2}, \cdots, N_{o, j_n})+1
\label{horton2}
\end{equation}
where, $N=\{N_{o,j_1},N_{o,j_2},\cdots N_{o,j_n}\}$ is the order value of sub-dendrites of the $j$-$th$ dendrite in set  $O$, $j_n$ is $n$-$th$ sub-dendrite of the $j$-$th$ dendrite, and $n$ is the number of the sub-dendrites of the $j$-$th$ dendrite.

\subsubsection{Shreve scheme} In this coding scheme, the morphological structure is quantitatively analyzed by using the adaptive Shreve encoding scheme~\cite{Shreve1967Infinite}. Terminal dendrites of set $E$ are given by the order value as 1, $N_{e,i} =1$. The non-terminal dendrites order values are given as:
\begin{equation}
N_{o,j} =\sum_{k=1}^{n} N_{o, j_k}
\label{shreve2}
\end{equation}
where $N_{o, j_k}$ is the order values of the $k$-$th$ child dendrite of the $j$-$th$ in set  $O$, and $n$ is the number of the child dendrites of the $j$-$th$ dendrite. 

\subsubsection{Elect scheme}
In this scheme, different dendritic geometries result in different electricity properties of dendrites, so one can use the feature of input resistance to analyze tree dendrites, so that the values of the input resistance of dendrites are used as the dendritic order values:
\begin{equation}
N = (N_{e,i} , N_{o, j}) = \textrm{Input resistance}
\end{equation}
In this model, different functional regions of PCs can be characterized by their order values as 
\begin{equation}
dend\in \left\{ \begin{array}{ll}
Spiny, & \textrm{if }\quad N \geq {s2} \\
Smooth, & \textrm{if}\quad  s1<N<s2\\
Trunk ,& \textrm{if}\quad N\leq{s1}
\end{array} \right.
\label{classify1}
\end{equation}
Note Eq.~\eqref{classify1} is operating in an opposite way as Eq.~\eqref{classify}, since for Branch, Horton and Shreve methods, tree order values are increased from spiny to trunk, but Elect method is working as a decreasing process.

\begin{table}[ptb]
	\renewcommand{\arraystretch}{1}
	\caption{Best thresholds of reduction models for each cell.}
	\begin{tabular}{|p{0.9cm}<{\centering}|p{0.3cm}<{\centering}|p{1.7cm}<{\centering}|p{1.7cm}<{\centering}|p{1.7cm}<{\centering}|p{1cm}<{\centering}|p{1cm}<{\centering}|p{1cm}<{\centering}|p{1cm}<{\centering}|p{1cm}<{\centering}|p{0.5cm}<{\centering}|p{0.5cm}<{\centering}|p{0.5cm}<{\centering}|}
		\hline

       & &\textbf{Guinea-pig1}& \textbf{Guinea-pig2} & \textbf{Guinea-pig3}&\textbf{Mouse1}& \textbf{Mouse2 }& \textbf{Mouse3 }& \textbf{Mouse4 }&\textbf{ Rat1} & \textbf{Rat2 }& \textbf{Rat3 } \\
        \hline
       \multirow{2}{*}{\textbf{Branch}}&\textbf{s1} &3&3&3&3&3&3&3&3&3&3\\
       \cline{2-12}
      & \textbf{s2} &8&8&8&8&8&8&8&8&8&8\\
           \hline
       \multirow{2}{*}{\textbf{Horton}}&\textbf{s1} &3&3&3&3&3&3&3&3&3&3\\
       \cline{2-12}
      & \textbf{s2} &30&90&20&50&50&30&30&20&30&20\\
           \hline
       \multirow{2}{*}{\textbf{Elect}}&\textbf{s1} &13&16&18&62&87&35&69&33&27&86\\
       \cline{2-12}
      & \textbf{s2} &17&17&19&63&88&36&70&34&29&87\\
      \hline
      \multirow{2}{*}{\textbf{Shreve}}&\textbf{s1} &10&10&10&10&10&10&10&10&10&10\\
       \cline{2-12}
      & \textbf{s2} &30&30&30&30&30&30&30&30&30&30\\
		\hline
	\end{tabular}
	\label{best_threshold}
\end{table}

Once the coding stage is finished, the rest of the work-flow process of reducing neuronal morphology is illustrated in Fig.~\ref{cluster} (C). One can collect those dendrites according to their order values into a cluster, such that there are three sets of clusters as $C_{trunk}$ for trunk dendrites, $C_{smooth}$ for smooth dendrites, and $C_{spiny}$ for spiny dendrites. Within each set, there is a series of subsets of the cluster in each region of the dendritic field.
Such a subset of clusters can be mapped into one single compartment by using the same merging rule as in Ref.~\cite{Marasco2013Using}, then ionic and synaptic conductances are scaled by a factor to preserve membrane area~\cite{Marasco2013Using} or preserve volume. 
To preserve membrane area in Horton, Shreve, and Elect models, we use a factor $f_{s,j}^{eq}=\frac{\sum\nolimits_js_j}{s_j^{eq}},
$ where $\sum\nolimits_js_j$ is the sum of the membrane area of each section in the clusters, and $s_j^{eq}$ is the membrane area of the equivalent compartment. However, in Branch model, one has to preserve volume instead of membrane area, we use a factor $f_{v,j}^{eq}=\frac{\sum\nolimits_jv_j}{v_j^{eq}}$, where $\sum\nolimits_jv_j$ is the sum of the volumes of each section in the clusters, and $v_j^{eq}$ is the volume of the equivalent compartment. Finally, synapse locations are sequentially mapped in each reduced models in the same way as in Ref.~\cite{Marasco2013Using}.

\subsection{Multi-compartment model of Purkinje cell }

To compare different reduced models with full morphology, we used multi-compartment models based on ten detailed morphological 3D reconstructions of Purkinje cells of guinea-pig, mouse, and rat, from public archive \textit{neuromorpho.org}. Specific capacitance was set to $0.8F/{cm}^2$ in the soma, and $1.5F/{cm}^2$ in trunk dendrites, smooth dendrites, and spiny dendrites. Internal axial resistivity was set to $250 \Omega/cm$ similar to the values used in Ref.~\cite{Ostojic2015Neuronal,Rapp1994Physiology}.

The same set of parameter values of passive properties, such as voltage-dependent ionic channels, kinetic, and distribution, was used for all morphologies of PCs. There are thirteen different types of voltage-gated ion channels modeled, eight of which (P-type $Ca^{2+}$ channel, T-type $Ca^{2+}$ channel, class-E $Ca^{2+}$ channel, persistent $K^+$ channel, A-type $K^+$ channel, D-type $K^+$ channel, delayed rectifier, decay of sub-membrane $Ca^{2+}$) were inserted into soma and dendrites. In addition, three ion channels (fast and persistent sodium channel, anomalous rectifier channel) were solely added to soma, and two ion channels (high-threshold calcium-activated potassium channel, low-threshold calcium-activated potassium channel) were solely added to dendrites~\cite{De1994An,Miyasho2001Low}.

\subsection{ Stimulation protocols} 

Stimulation of PC was implemented by parallel fiber (PF) inputs, where synaptic input from each PF to PC is characterized by AMPA receptors~\cite{Gao2012Distributed}. Following the typical values estimated from experiments, a total of 1000 PF connections for a single PC was used as previously~\cite{Masoli2017Synaptic}. AMPA synapses were inserted only in the spiny dendrites with a random distribution. AMPA synapses were modeled as a double exponential conductance change with 0.5 ms and 1.2 ms for rising and decay time respectively~\cite{De1994An1}, and the maximal synaptic conductance was drawn from a Gaussian distribution as $\rm{5\pm{0.5 }}$ nS.  

In addition, PCs are affected by direct synaptic inhibition coming from molecular layer interneurons, we used a total of 500 inhibitory connections on a single PC. These synapses were randomly distributed on the spiny dendrites, and modeled as $\rm GABA_{A}$~\cite{he2015interneuron} with a double exponential conductance change with 0.5 ms and 2.5 ms for rising and decay time respectively, and the maximal synaptic conductance was drawn from a Gaussian distribution as $\rm 5\pm 0.5$ nS.

PCs aligned on the mediolateral axis receive about 175,000 PF inputs~\cite{Napper1988Number,Hoxha2016Modulation}. This arrangement results in the hypothesis that the evoked PC firing for temporal control of movement is encoded in the cerebellum by beams of synchronously active PCs~\cite{Jaeger2003No}. So in our model, PCs response were driven by 1000 synchronized PF inputs random distributed on spiny dendrites~\cite{Hoxha2016Modulation,Su2012Target}, i.e., simulations were run with synchronized stimulation protocol. 

A single stimulation consists of a sequence of spikes containing spike times and inter-spike intervals (ISIs), so we can generate a successive spike train by the previous spike plus regular or irregular time intervals. There are three types of stimulation used in this study depending on the sampling process.

\subsubsection{Poisson process}

Spike trains were modeled using a homogeneous Poisson process in which the ISI distribution is exponential. The probability density function of ISI $\tau$ is given by:
\begin{equation}
p(\tau)=re^{-r\tau}
\end{equation}
where r is mean firing rate. The mean $<\tau>$ and standard deviation $\sigma$ of ISIs are both $1/r$.

\subsubsection{Renewal process}
A simple way to generate a spike train based on renewal process is to start with a Poisson spike train and delete all but every $k$-th spike. In this way, a spike train is obtained with ISIs $\tau$ given by gamma probability density function:
\begin{equation}
p( \tau )=(k\tau)^k\tau^{k-1}e^{-kr\tau}/{(k-1)!}
\end{equation}
where $k$ is the order of gamma distribution, and $r$ is mean firing rate. The mean $<\tau>$ and standard deviation $\sigma$ are $1/r$ and$<\tau> /\sqrt{k}$, respectively. Here k is set to 2 through the whole study. 

\subsubsection{Modulated renewal process}

Recent work on the statistical modeling of neural responses has focused on modulated renewal processes in which the spike rate is a function of the recent spiking history. Here we modeled the modulation of the firing rate $r(t)$ as a synaptic input frequency such that  
\begin{equation}
r(t)=Asin(2\pi{ft}+\theta)+A
\end{equation}
where the modulation of the firing rate is thus fully specified by its amplitude $A$ and frequency $f$ of the sinusoidal component. Thus one can generate a spike train that satisfies this firing rate $ r(t)$ by the time-rescaling method~\cite{Brown2002The,Pillow2009Time, Zampini2016Mechanisms}. 

\subsection{Data analysis}

The full and reduced PC morphology models were stimulated in NEURON 7.4. Data analysis was implemented with MATLAB. Unless otherwise noted, all simulations run 2000 ms. where data from the last 1000 ms simulation were used to analyze the results. The time step 0.025 ms was used for all simulations. For analysis, we used four measures to characterize the performance of the reduced method: accuracy, simplification, efficiency and spike shape change. 

\subsubsection{Accuracy}
The main purpose of a reduced model is to maintain the input-output (I/O) property on the basis of simplification of neuronal morphology. We evaluated the accuracy of the I/O property by comparing the spike trains generated before and after reduction~\cite{Marasco2013Using}:
\begin{equation}
\rm \textrm{Accuracy}=\frac{TP+TN}{TP+TN+FP+FN}
\end{equation}
where TP (True Positives) is the number of spikes from the full model that are also found in the reduced model.  TN (True Negatives) is the number of intervals that the neuron does not fire in both the full and reduced models. FP (False Positives) is the number of mismatched spikes in the reduced models. FN (False Negatives) is the number of spikes from the full model that are not matched in the reduced models.

\subsubsection{Simplification}
Morphological simplification is the most basic requirement for a reduced method. 
The full morphological structure can be seen as being made up by a series of cylindrical segments with different lengths and diameters.
Therefore, we characterized the degree of simplification as the ratio of the segments of morphological structure before and after reduction:
\begin{equation}
\rm Simplification = \frac{SEG_{\textrm{full}}-SEG_{\textrm{reduced}}}{SEG_{\textrm{full}}}
\end{equation}
where $\rm SEG_{\textrm{reduced}}$ is the number of the segments in reduced morphological structure, and $\rm SEG_{\textrm{full}}$ is the number of the segments in full morphological structure.

\subsubsection{Efficiency}
Another feature of a reduced model is to improve computing efficiency, as one wants to compute the neuronal dynamics as fast as one can, in particular in large-scale network simulation. We evaluated this feature by the ratio between the runtime of the full model and reduced model in the same computing environment: 
\begin{equation}
\rm Efficiency = \frac{ Runtime_{full}}{Runtime_{reduce}}
\end{equation}
where $\rm Runtime_{full}$ is the run time of the full model and $\rm Runtime_{reduce}$  is the run time of the reduced model.

\subsubsection{Spike shape}
In addition, we also analysis the accuracy of spike shape including spike width and spike amplitude. We evaluated this feature by comparing the changes in the full model and the reduce model：
\begin{equation}
\Delta Change_X=|Full_X-Reduce_X|
\end{equation} Where X represents spike width or spike amplitude, $Full_X$ and $Reduce_X$ are the mean spike width or spike amplitude in spike trains of the full model and reduced model, respectively.

\section{Results}

We consider four different reduction schemes to simplify the morphology of PCs from three species, then study their effects on PC responses to synaptic inputs. To investigate how the coding capacity of PCs is affected by morphological structure reductions, we evaluate the performance of the reduced model with four measures: simplification, accuracy, efficiency, and spike shape change. Furthermore, the coding capacity of PCs, in terms of firing rate coding, spike timing coding, and modulated firing amplitude and phase under different input synaptic frequencies are studied as well. 

\subsection{Performance evaluation of morphology reduction}

\begin{figure*}[pht]	
	\centering	
    \includegraphics[width=\columnwidth]{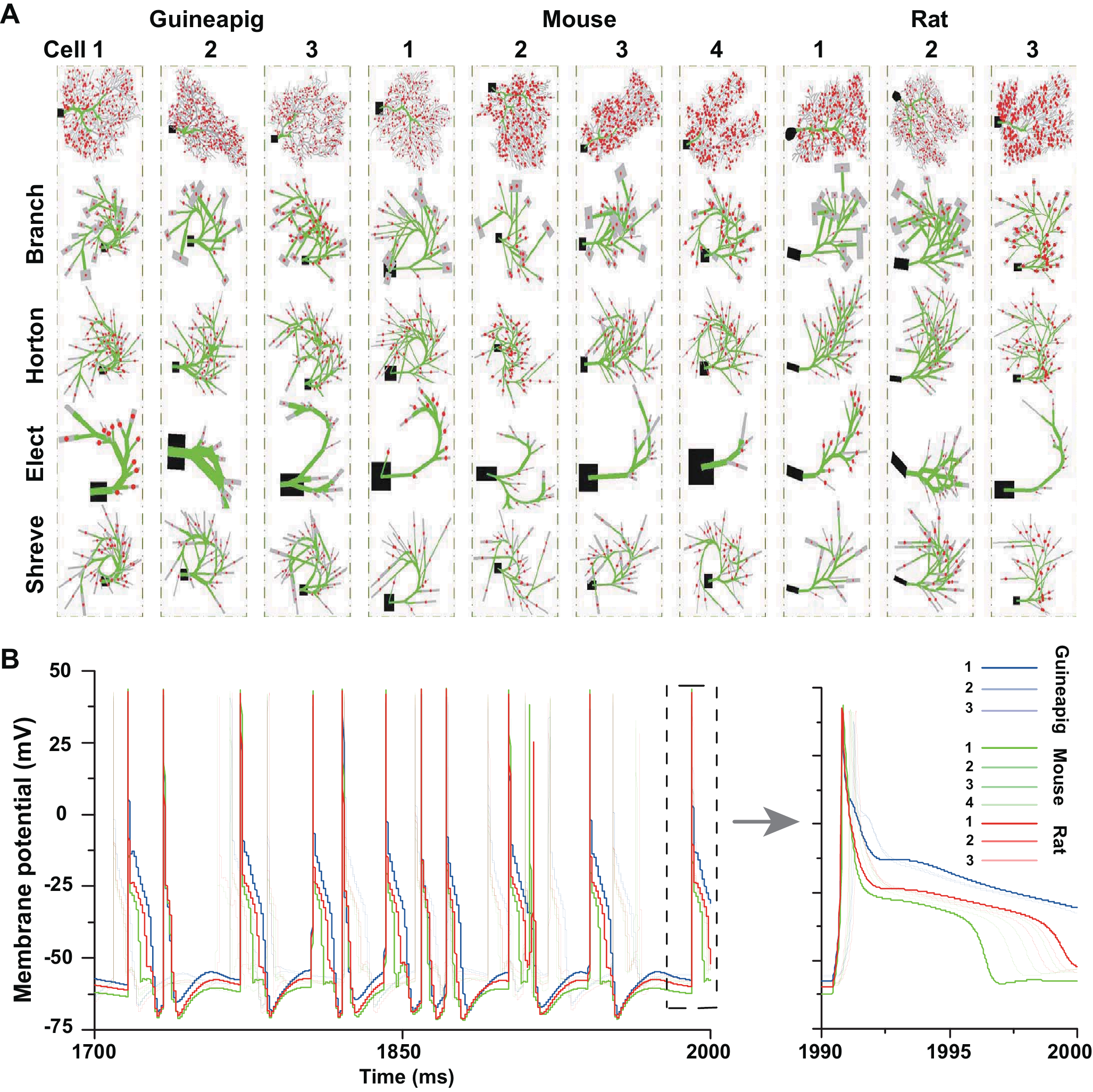}
	\caption{
    Detailed and simplified PC morphologies. (A) Totally ten PCs from three species are reduced by four different schemes, Branch, Horton, Elect and Shreve methods. Spiny dendrites in gray, smooth dendrites and initial major branches in green. Spiny dendrites receives 1000 excitatory AMPA-type synapses from parallel fibers (red dots). (B) Hyperpolarization phases after spiking are different in ten full PC morphologies. Poisson stimulation at 50Hz. 
    }
    \label{ten-morphology}
\end{figure*}

Ten different PCs with different morphologies from guinea-pig, mouse, and rat were studied. The reduced morphologies were generated by four types of reduction schemes for each cell as shown in Fig.~\ref{ten-morphology} (A). Four reducing schemes show different simplifications for a particular morphology. However, there is a large variety of neuronal morphology for each cell from each species. Although the same set of parameters were used in all of ten PCs, different morphologies result in different dynamics of hyperpolarization. Such a difference is larger across species and smaller within the same species (Fig.~\ref{ten-morphology} (B)). 

\begin{figure*}[tpht]	
	\centering	
    \includegraphics[width=\columnwidth]{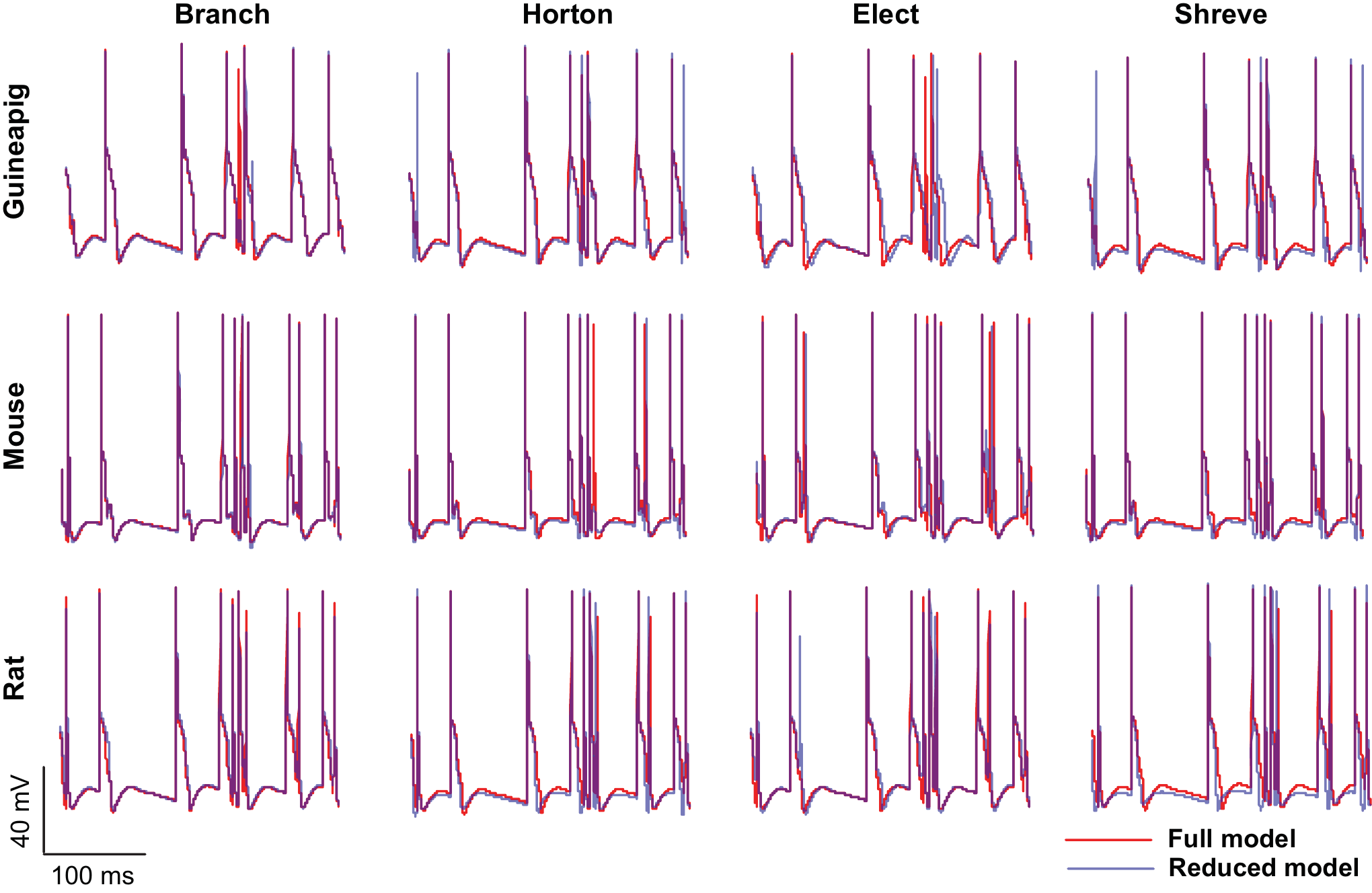}
	\caption{ Spiking dynamics of full and reduced PC models. Membrane potential traces recorded from soma of three example PCs of guinea-pig, mouse and rat with full model (red) and four reduced models (blue). Poisson stimulation at 50Hz. 
    \label{membrane}
    }
\end{figure*}

Then we focus on three example PCs, guinea-pig1 (v\_e\_purk1Mod), mouse1 (e4cb2a2Mod), and rat1 (p20) to illustrate the spiking dynamics of full and reduced morphology. Throughout the study, these three example PCs are used to represent three species respectively. In general, for all PCs, the spiking dynamics of reduced morphology in all four schemes matches that of the full model very well (Fig.~\ref{membrane}).

\begin{table}[tpb]
	\renewcommand{\arraystretch}{1}
	\caption{Accuracy of reduced models. Values are mean $\pm$ STD calculated from 21  sets of Poisson stimulation from 10 to 1K Hz. The best reduced method for each cell is marked in bold.}
	\label{Table accuracy}
	\begin{tabular}{|p{1cm}<{\centering}|p{1.7cm}<{\centering}|p{1.7cm}<{\centering}|p{1.7cm}<{\centering}|p{1cm}<{\centering}|p{1cm}<{\centering}|p{1cm}<{\centering}|p{1cm}<{\centering}|p{0.9cm}<{\centering}|p{0.9cm}<{\centering}|p{0.9cm}<{\centering}|}
		\hline
        &\textbf{Guinea-pig1}& \textbf{Guinea-pig2} & \textbf{Guinea-pig3}&\textbf{Mouse1}& \textbf{Mouse2 }& \textbf{Mouse3 }& \textbf{Mouse4 }&\textbf{ Rat1} &\textbf{  Rat2} &\textbf{ Rat3}\\
        \hline
       \textbf{Branch}&\boldmath{$0.976\pm{0.012}$}&$0.940\pm{0.031}$ &\boldmath{$0.973\pm{0.017}$} &$0.926\pm{0.057}$ &         $0.905\pm{0.063}$&\boldmath{$0.984\pm{0.013}$}&\boldmath{$0.965\pm{0.019}$ }&$0.941\pm{0.036}$&\boldmath{$0.963\pm{0.024}$} &\boldmath{$0.964\pm{0.029}$}\\
       	\hline	
         \textbf{Horton} & $0.954\pm{0.021}$&\boldmath{$0.942\pm{0.029}$ }& $0.971\pm{0.015}$ &$0.860\pm{0.078}$&$0.885\pm{0.064}$ & $0.949\pm{0.032}$& $0.880\pm{0.069}$ & $0.934\pm{0.038}$&$0.926\pm{0.035}$ &$0.955\pm{0.028}$ \\
         \hline	
          \textbf{Elect}& $0.886\pm{0.021}$& $0.825\pm{0.048}$&$0.880\pm{0.040}$&\boldmath{$0.970\pm{0.019}$}&$0.860\pm{0.063}$&$0.938\pm{0.029}$& $0.848\pm{0.047}$& \boldmath{$0.968\pm{0.015}$}&$0.781\pm{0.074}$&$0.850\pm{0.052}$\\
          	\hline	
            \textbf{Shreve}& $0.943\pm{0.023}$&$0.924\pm{0.035}$&$0.972\pm{0.012}$& $0.906\pm{0.054}$&\boldmath{$0.913\pm{0.052}$}&$0.959\pm{0.020}$&$0.939\pm{0.033}$&$0.889\pm{0.064}$& $0.954\pm{0.017}$&$0.921\pm{0.050}$\\
		\hline
            \textbf{Marosco}& $0.932\pm{0.029}$&$0.929\pm{0.029}$&$0.930\pm{0.027}$& $0.872\pm{0.092}$&$0.881\pm{0.059}$&$0.955\pm{0.022}$&$0.939\pm{0.033}$&$0.906\pm{0.051}$& $0.947\pm{0.019}$&$0.908\pm{0.057}$\\
		\hline
	\end{tabular}
\end{table}

\begin{table}[htb]
	\renewcommand{\arraystretch}{1}
	\caption{Simplification  of reduced models.}
	\label{Table simplification}
	\begin{tabular}{|p{1cm}<{\centering}|p{1.7cm}<{\centering}|p{1.7cm}<{\centering}|p{1.7cm}<{\centering}|p{1cm}<{\centering}|p{1cm}<{\centering}|p{1cm}<{\centering}|p{1cm}<{\centering}|p{0.9cm}<{\centering}|p{0.9cm}<{\centering}|p{0.9cm}<{\centering}|}
		\hline
        &\textbf{Guinea-pig1}& \textbf{Guinea-pig2} & \textbf{Guinea-pig3}&\textbf{Mouse1}& \textbf{Mouse2 }& \textbf{Mouse3 }& \textbf{Mouse4 }&\textbf{ Rat1} &\textbf{  Rat2} &\textbf{ Rat3}\\
        \hline
       \textbf{Branch}&85.5\%&93.2\%&80.1\%&93.3\%&90.3\%&86.0\%&85.7\%&86.6\%&90.7\%&82.8\%\\
       	\hline	
         \textbf{Horton} &79.5\%&82.1\%&79.5\%&79.7\%&77.2\%&78.5\%&78.8\%&81.8\%&83.6\%&80.6\% \\
         \hline	
          \textbf{Elect}&\textbf{96.7\%}&\textbf{96.9\%}& \textbf{96.2\%}&\textbf{95.7\%}&91.9\%&\textbf{97.1\%}&\textbf{98.9\%}&93.1\%&\textbf{94.3\%}&\textbf{97.7\%}\\
          	\hline	
            \textbf{Shreve}& 88.8\%&90.7\%&87.6\%&91.7\%&88.8\%&89.7\%&88.5\%&93.3\%&91.7\%&91.0\%\\
		\hline
              \textbf{Marosco}& 91.4\%&93.0\%&95.1\%&93.9\%&\textbf{92.2}\%&92.7\%&95.1\%&\textbf{94.6\%}&94.1\%&93.9\%\\
		\hline
	\end{tabular}
\end{table}

\begin{table}[tb]
	\renewcommand{\arraystretch}{1}
	\caption{Efficiency of reduced models. Values are mean $\pm$ STD. }
	\label{Table Efficiency }
	\begin{tabular}{|p{0.9cm}<{\centering}|p{1.7cm}<{\centering}|p{1.7cm}<{\centering}|p{1.7cm}<{\centering}|p{1cm}<{\centering}|p{1cm}<{\centering}|p{1cm}<{\centering}|p{1cm}<{\centering}|p{0.9cm}<{\centering}|p{1.1cm}<{\centering}|p{1.1cm}<{\centering}|}
		\hline
        &\textbf{Guinea-pig1}& \textbf{Guinea-pig2} & \textbf{Guinea-pig3}&\textbf{Mouse1}& \textbf{Mouse2 }& \textbf{Mouse3 }& \textbf{Mouse4 }&\textbf{ Rat1} &\textbf{ Rat2} &\textbf{ Rat3}\\
        \hline
       \textbf{Branch}&$19\pm3$&$22\pm8$&$9\pm2$&$24\pm7$&$20\pm2$&$15\pm2$&$15\pm3$&$13\pm1$&$21\pm6$&$12\pm3$\\
       	\hline	
         \textbf{Horton} &$9\pm2$&$9\pm2$&$8\pm1$&$11\pm1$& $9\pm1$&$9\pm1$&$10\pm2$&$11\pm2$&$13\pm2$&$10\pm2$ \\
         \hline	
          \textbf{Elect}&\boldmath{$40\pm8$}&\boldmath{$59\pm12$}&\boldmath{$46\pm8$}& \boldmath{$56\pm9$}&\boldmath{$30\pm4$}&\boldmath{$64\pm6$}&\boldmath{$72\pm1$}&$16\pm1$&\boldmath{$39\pm13$}&\boldmath{$41\pm2$}\\
          	\hline	
            \textbf{Shreve}&$16\pm3$&$18\pm3$&$12\pm1$&$18\pm3$&$13\pm1$&$14\pm1$& $14\pm2$&$20\pm2$&$17\pm5$&$16\pm4$\\
                    	\hline	
            \textbf{Marosco}&$24\pm6$&$27\pm4$&$35\pm4$&$31\pm3$&$23\pm2$&$24\pm3$& $26\pm1$&\boldmath{$22\pm1$}&$24\pm4$&$22\pm3$\\
		\hline
	\end{tabular}
\end{table}

\begin{table}[tpb]
	\renewcommand{\arraystretch}{1}
	\caption{  Change of spike amplitude in reduced models (mv, mean $\pm$ STD).  }
	\label{Table amplitude}
	\begin{tabular}{|p{0.9cm}<{\centering}|p{1.7cm}<{\centering}|p{1.7cm}<{\centering}|p{1.7cm}<{\centering}|p{1cm}<{\centering}|p{1cm}<{\centering}|p{1cm}<{\centering}|p{1cm}<{\centering}|p{0.9cm}<{\centering}|p{0.9cm}<{\centering}|p{0.9cm}<{\centering}|}
		\hline
        &\textbf{Guinea-pig1}& \textbf{Guinea-pig2} & \textbf{Guinea-pig3}&\textbf{Mouse1}& \textbf{Mouse2 }& \textbf{Mouse3 }& \textbf{Mouse4 }&\textbf{ Rat1} &\textbf{  Rat2} &\textbf{ Rat3}\\
        \hline
       \textbf{Branch}&\boldmath{$1.08\pm{0.53}$}&$3.63\pm{1.95}$ &$1.63\pm{0.98}$ &$0.68\pm{0.88}$ &         $6.23\pm{2.74}$&\boldmath{$0.48\pm{0.17}$}&$1.03\pm{0.57}$ &$3.71\pm{1.82}$&$3.87\pm{1.87}$&\boldmath{$0.78\pm{0.27}$}\\
       \hline
         \textbf{Horton} & $2.08\pm{1.02}$&$3.12\pm{1.37}$ &$1.42\pm{0.62}$ &$1.03\pm{0.24}$&$7.30\pm{3.38}$ & $0.63\pm{0.16}$& \boldmath{$0.55\pm{0.58}$} & $1.81\pm{0.55}$&$0.62\pm{0.27}$ &$0.91\pm{0.28}$ \\
         \hline	
          \textbf{Elect}& $1.58\pm{0.80}$& \boldmath{$3.09\pm{1.36}$}& \boldmath{$1.24\pm{1.00}$}&\boldmath{$0.54\pm{0.22}$}&$12.10\pm{5.89}$&$2.00\pm{0.74}$& $5.17\pm{3.30}$& \boldmath{$0.95\pm{0.50}$}&$2.85\pm{0.70}$&$1.72\pm{1.43}$\\
          	\hline	
            \textbf{Shreve}& $3.02\pm{1.40}$&$4.98\pm{2.60}$&$1.95\pm{1.02}$& $0.86\pm{0.19}$&\boldmath{$4.95\pm{2.49}$}&$0.81\pm{0.26}$&$0.89\pm{0.65}$&$3.56\pm{1.32}$& \boldmath{$0.56\pm{0.18}$}&$1.98\pm{0.53}$\\
		\hline
            \textbf{Marosco}& $3.86\pm{1.88}$&$4.40\pm{2.18}$&$4.00\pm{1.83}$& $1.80\pm{0.48}$&$6.61\pm{2.89}$&$1.04\pm{0.30}$&$0.80\pm{0.61}$&$3.81\pm{1.34}$& $0.79\pm{0.24}$&$2.55\pm{1.24}$\\
		\hline
	\end{tabular}
\end{table}

\begin{table}[tpb]
	\renewcommand{\arraystretch}{1}
	\caption{Change of spike width in reduced models (ms, mean $\pm$ STD). } 
	\label{Table width}
	\begin{tabular}{|p{0.9cm}<{\centering}|p{1.7cm}<{\centering}|p{1.7cm}<{\centering}|p{1.7cm}<{\centering}|p{1cm}<{\centering}|p{1cm}<{\centering}|p{1cm}<{\centering}|p{1cm}<{\centering}|p{0.9cm}<{\centering}|p{0.9cm}<{\centering}|p{0.9cm}<{\centering}|}
		\hline
        &\textbf{Guinea-pig1}& \textbf{Guinea-pig2} & \textbf{Guinea-pig3}&\textbf{Mouse1}& \textbf{Mouse2 }& \textbf{Mouse3 }& \textbf{Mouse4 }&\textbf{ Rat1} &\textbf{  Rat2} &\textbf{ Rat3}\\
        \hline
       \textbf{Branch}&\boldmath{$0.29\pm{0.23}$}&$0.73\pm{0.29}$ &\boldmath{$0.18\pm{0.16}$} &$0.20\pm{0.25}$ &       $1.02\pm{0.88}$&\boldmath{$0.06\pm{0.08}$}&$0.53\pm{0.50}$ &$0.28\pm{0.26}$&$0.92\pm{1.07}$&\boldmath{$0.09\pm{0.09}$}\\
       \hline
         \textbf{Horton} & $0.34\pm{0.27}$&\boldmath{$0.41\pm{0.32}$} &$0.23\pm{0.15}$  &\boldmath{$0.20\pm{0.18}$}&$0.84\pm{0.44}$ & $0.25\pm{0.27}$& $0.81\pm{0.32}$ & $0.18\pm{0.19}$&$0.30\pm{0.19}$&$0.13\pm{0.18}$\\
         \hline	
          \textbf{Elect}& $0.81\pm{0.62}$& $1.39\pm{1.78}$& $0.44\pm{0.45}$&$0.26\pm{0.14}$&\boldmath{$0.51\pm{0.74}$}&$0.34\pm{0.21}$& $0.4\pm{0.4}$&\boldmath{ $0.11\pm{0.12}$}&$0.30\pm{0.23}$&$0.49\pm{0.19}$\\
          	\hline	
            \textbf{Shreve}& $0.42\pm{0.39}$&$0.59\pm{0.42}$&$0.27\pm{0.35}$& $0.23\pm{0.14}$&$0.92\pm{0.45}$&$0.21\pm{0.15}$&$0.55\pm{0.32}$&\boldmath{$0.37\pm{0.37}$}& $0.28\pm{0.22}$&$0.36\pm{0.43}$\\
           \hline
            \textbf{Marosco}& $0.49\pm{0.33}$&$0.52\pm{0.33}$&$0.41\pm{0.32}$& $0.31\pm{0.16}$&$0.57\pm{0.41}$&$0.18\pm{0.23}$&\boldmath{$0.31\pm{0.13}$}&$0.34\pm{0.46}$& $0.32\pm{0.22}$&$0.47\pm{0.69}$\\
		\hline
	\end{tabular}
\end{table}

It is especially important to ensure the firing accuracy of the reduced model, which can be seen in Fig.~\ref{membrane}. The detailed measure quantified by \textit{accuracy} shown in Table~\ref{Table accuracy}, which describes how a somatic membrane potential trace of the full model can be reproduced in different reduced models. For comparison, Marosco model was included~\cite{Marasco2013Using}. The accuracy is relatively diverse in four reduced schemes, but in general, Branch model has the highest accuracy. However, Marosco reduced model does not archive accurate firing activities.

The second measure is \textit{simplification} that characterizes how the morphologies are changed by reduced models. Reduced model simplifies complex tree structures into a much smaller set of dendrites that can be quantified by the degree of simplification. Table~\ref{Table simplification} shows that the Elect model has the largest degree of simplification such that the reduced morphology is most compact. In contrast, Horton model has a minimum degree of simplification. Marosco model has a slightly larger degree of simplification than Elect model for mouse2 and rat1.

A direct outcome of simplification is that the simplest model needs less computing time for simulation. It is feasible to ensure the application of reduced models on large-scale networks only by improving computational efficiency while high accuracy is maintained. Table~\ref{Table Efficiency } shows the quantification of \textit{runtime}, where Elect model achieves the most efficient computation as it has the highest degree of simplification, except that Marosco model is fastest for rat1.

Furthermore, it is also important to maintain the accuracy of spike shape, which can be measured by the change of spike amplitude in Table~\ref{Table amplitude} and spike width in  Table~\ref{Table width}, where Branch model has higher accuracy for more cells. 

Given that the accuracy as the most important factor for reduction, Branch model is the best one for the performance of accuracy of spike rate, the degree of simplification, runtime and spike shape, are also good. The reason for this is mainly due to the fact that the scaling factor in the simplification process is to preserve the cell volume in Branch model, but the other three models preserve the cell membrane area. Thus, we conclude that cell volume may play an important role in shaping Purkinje cell firing.

\subsection{ PC firing rate coding}

\begin{figure*}[thtbp]	
	\centering
		\includegraphics[width=\columnwidth]{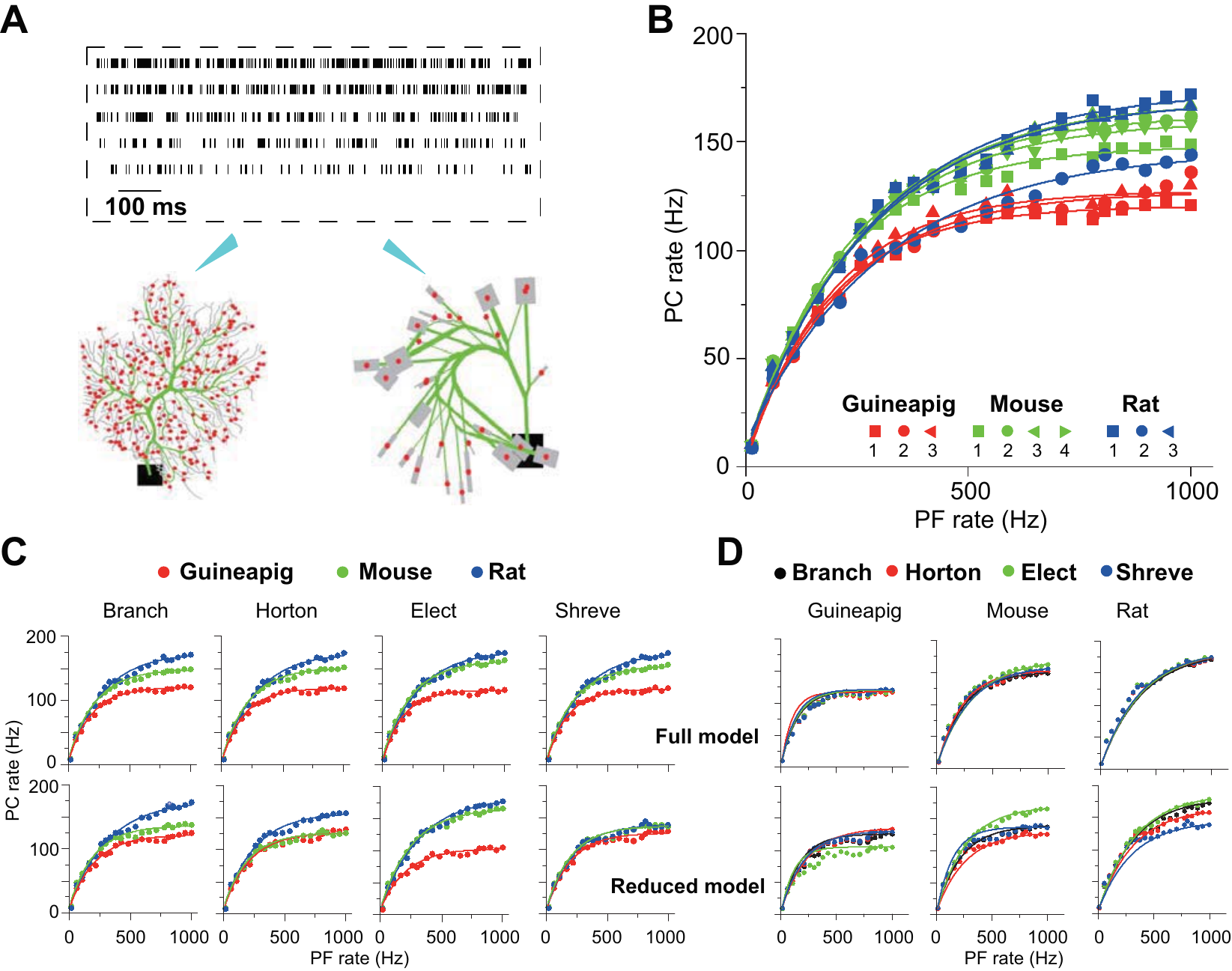}
	\caption{ PC firing responses in full and reduced models. (A) Schematic view of Poisson stimulation sequences from 60 Hz to 260 Hz injected to spiny dendrites of full (left) and reduced (right) models. (B) Firing response curves of ten PCs with full morphology. (C) and (D) Comparison of firing response curves of three example PCs from guinea-pig, mouse and rat under four reduction schemes, Branch, Horton, Elect and Shreve, respectively. PC curves are grouped by species in (C) and by reduction schemes in (D). Note that there are four PC response curves in one full model, since each is a realization of random distribution of PF input synapses. Solid curves in (B-D) are fitted exponential functions. Poisson stimulation in all cases.
    }
    \label{tenpoissonfre}
\end{figure*}

To characterize the performance of reduced models, we carry out a wide range of stimulation frequencies from 10Hz to 1KHz sampled in Poisson process as illustrated in Fig.~\ref{tenpoissonfre} (A), which is in the same range as observed \textit{in vivo} experiments where granule cells can discharge from a few Hz up to 1K Hz~\cite{Valera2012Adaptation}. 

The cerebellum can control high-precise motor patterns with millisecond resolution by using a wide range of action potential firing rates~\cite{Ostojic2015Neuronal,Amir2011Cerebellum,Jelitai2016Dendritic}.
Fig.~\ref{tenpoissonfre} (B) shows ten PCs firing rate response curves under Poisson stimuli ranging from 10 Hz to 1K Hz in full morphology model. When such stimuli are used,  different species have their own response frequency range except for PC rat2 that has a lower range comparing to other two rat PCs. 

There is a variation of PC response curves within the same species due to their morphological differences. The PC firing rates dramatically increase with stimulations from 10 Hz to 180 Hz, and then slowly saturate at the higher stimulation frequencies, which is similar to the experimental observations found in awake animals where PCs exhibit an action potential rate between 30 and 200 Hz~\cite{Bryant2010Cerebellar,Cao2017Cerebellar}.  

By using the same three example PCs as in Fig.~\ref{membrane}, one can compare the detailed differences of PC response across different species and reduced models. Fig.~\ref{tenpoissonfre} (B) shows that guinea-pig PC response is significantly lower than mouse and rat, in particular, during high-frequency stimulation. This is consistent with the difference shown in the phase of hyperpolarization after spiking (Fig.~\ref{ten-morphology}(B)), even other parameters were fixed at the same.
This may imply that rat and mouse are able to react on millisecond timescale activities faster than guinea-pig.  
 
Four different reduced methods divide the PCs into different spiny dendrites, which leads to synaptic distribution locations slightly different, as a result, four corresponding full models are slightly different in their response curves (Fig.~\ref{tenpoissonfre} (C-D)). The response frequency has to be high enough when the stimulation frequency is high as evidenced by experimental data that PC could generate a high stage of firing rate to control fast timing patterns of the motor behavior. PC response curves can be well fitted by exponential function although the high frequency is slowly saturating.
 
After morphology is reduced, PC response curves are preserved from the full model, but there are some differences depending on the reduction schemes as shown in Fig.~\ref{tenpoissonfre} (C-D). Shreve and Horton models mediate PC firing rate decreased compared to full model at high stimulation frequencies. This could be seen from both mouse and rat, but guinea-pig shows slightly increased PC firing rate. In Elect model, guinea-pig shows a significant decrease in firing rate. In Branch model, the PC rate decreases in mouse while guinea-pig and rat show similar firing rate. The increased diversity at high frequency stimulation implies that complex spatial morphology structure can affect PCs to react to millisecond-scale activity.
In general, Branch model is the best model to preserve the input-output relationship of PCs in all species. 

The similar results were also observed when stimulation protocol is changed from Poisson to renewal process(Supplementary Fig.~1). Therefore, the PC firing responses vary depending on morphology and species rather than the types of stimulation used.

\subsection{PC timing coding}

\begin{figure*}[tbp]
    	\includegraphics[width=\columnwidth]{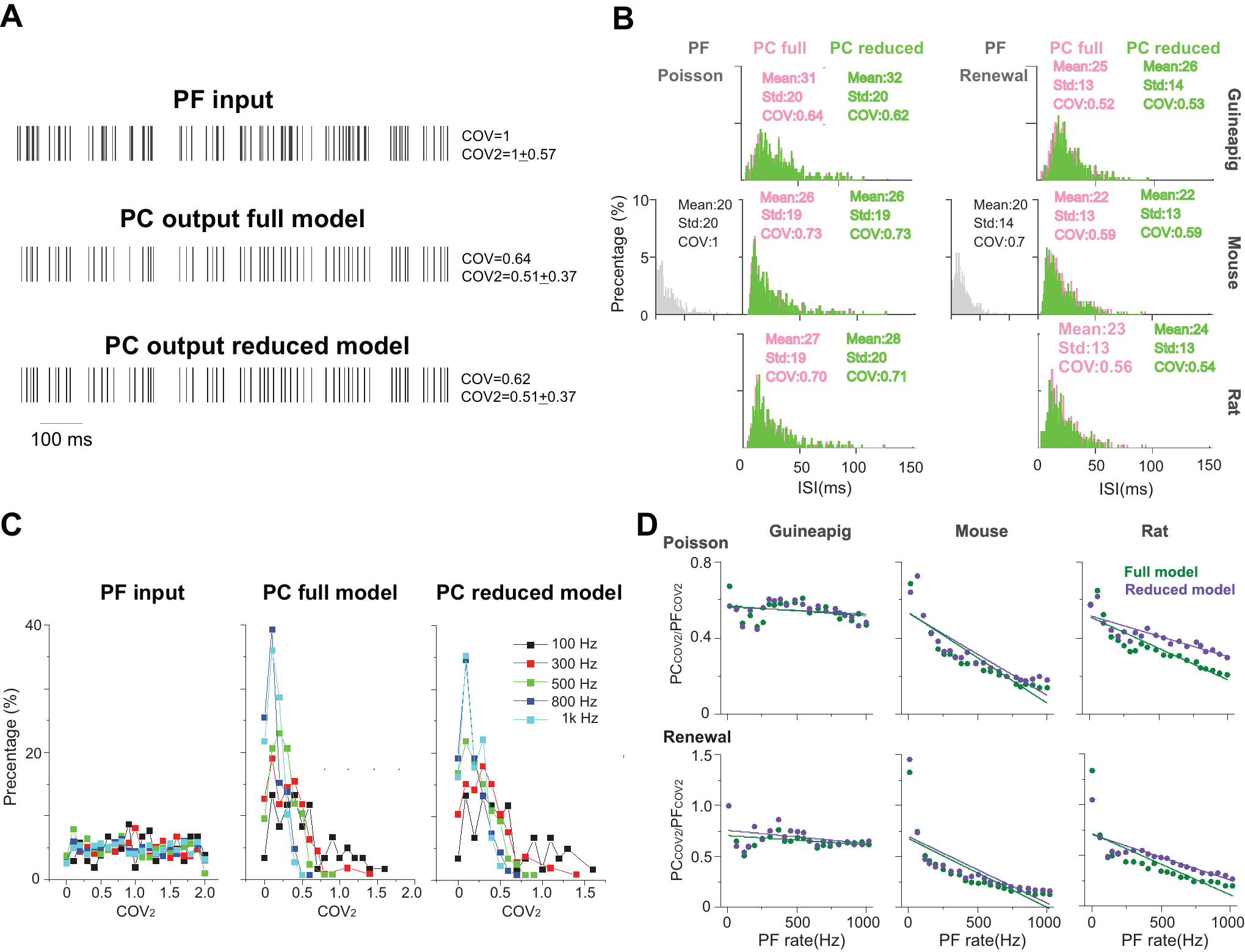} 
	\caption{ PC temporal responses under Poisson and renewal stimulation with Branch method.  
    (A) Schematic view of PF input (top) and PC output spike trains from full model (middle) and Branch model (bottom) from mouse. Poisson stimulation at 100Hz. 
    (B) ISI distribution of spike trains from PF (gray), PC full model (light red) and PC reduced model (green), respectively, under Poisson and renewal process stimulation for guinea-pig (top), mouse (middle), rat (bottom). All stimuli are at 50Hz for 10 seconds. Similarity between the ISI distributions of full and reduced model measured by p-value, Wilcoxon Rank-sum test. Guinea-pig, 0.51; Mouse, 0.97; Rat 0.26 for Poisson stimulation, and Guinea-pig, 0.63; Mouse 0.93; Rat, 0.32 for renewal stimulation.
    (C) Distribution of $\rm COV_2$ values obtained from spike trains of PF inputs (left), PC full model (middle), and Branch model (right) from mouse with Poisson stimulation of different frequencies. 
    (D) $\rm PC_{cov2}/PF_{cov2}$ showing the regularity between PF inputs and PC outputs for full model (green) and reduced model (purple) of guinea-pig, mouse and rat. Poisson and renewal process stimulation with different frequencies from 10 to 1K Hz. 
   }
	\label{ISIdistribution}
\end{figure*}

Purkinje cells transmit precise timing information to their downstream targets for precise control of motor-related tasks and conditioned behaviors~\cite{Ivry2004The,Koekkoek2003Cerebellar}. Here by using the full and reduced models with different stimulation protocols,  we investigate the effect of morphology structure on PC temporal coding, in particular, we focus on simple spikes that are a majority of PC spikes quantified by their inter spike intervals (ISIs), which has been shown statistics of ISIs play an important role in PC temporal coding~\cite{shin2006dynamic, Shin2007Regular}.

Fig.~\ref{ISIdistribution} shows the results of temporal coding precision of three same PCs from guinea-pig, mouse, and rat, under the stimulation protocols of Poisson and renewal processes. A schematic view of PF inputs and PC outputs of the full and reduced models illustrates that Branch model reduces the full morphology but keeps the temporal coding very accurately (Fig.~\ref{ISIdistribution}(A)). A common way to characterize the temporal structure of spike train is to use the coefficient of variation (COV) of ISIs. Based on this measure, the spike trains generated by the PCs are significantly more regular than those of PFs. The detailed statistics of ISIs of three example PCs with both full and reduced models confirm this observation (Fig.~\ref{ISIdistribution}(B)). Furthermore, the ISI distributions obtained by the reduced model are similar to those by the full model, which can be described by p-values from Wilcoxon Rank-sum test. For all three PCs and both stimulation protocols, p-values of two distributions of full and reduced model are non-significant. For the same frequency stimulus, the PC ISI distributions are different from each other across two stimulation protocols of Poisson and renewal process, and also different across species as well.   

Fine characterization of temporal precision of spike trains can be described by a modified measure $\rm COV_2=2|ISI_{n+1}-ISI_n|/(ISI_{n+1}+ISI_n)$, which can avoid that one of the ISI in the sequence affects global regularity~\cite{Hong2016Multiplexed,Shin2007Regular}. A single spike train can obtain a sequence of $\rm COV_2$ values (Supplementary Fig.3), and their distributions are shown in Fig.~\ref{ISIdistribution}(C) for several example frequencies of Poisson stimulation. Not surprisingly, the $\rm COV_2$ values of PF stimulation sequence are distributed uniformity with different Poisson frequencies. However, the $\rm COV_2$ values of PC spike trains have a wide distribution with the peak around 0.1.

One can average all $\rm COV_2$ values for each spike train, then use the value of the ratio $\rm PC_{COV_2}/PF_{COV_2}$ as an indicator to evaluate the regularity between PF inputs and PC outputs as shown in Fig.~\ref{ISIdistribution}(D), such that the ratio greater than 1 indicates that PF firing is more regular, otherwise, PC is more regular. Similar to ISI distribution, there are clear differences across species. Changing of firing regularity is found in mouse and rat, but not in guinea-pig.  It is worth noting that PC spike trains are more regular with the only exception that the ratio is close to 1 for renewal process in low stimulus frequency (10 Hz) (Supplementary Fig.~2(D-F)), while PC spike trains are always more regular for Poisson stimulation.  

\begin{figure*}[tbp]	
		\centering
		\includegraphics[width=\columnwidth]{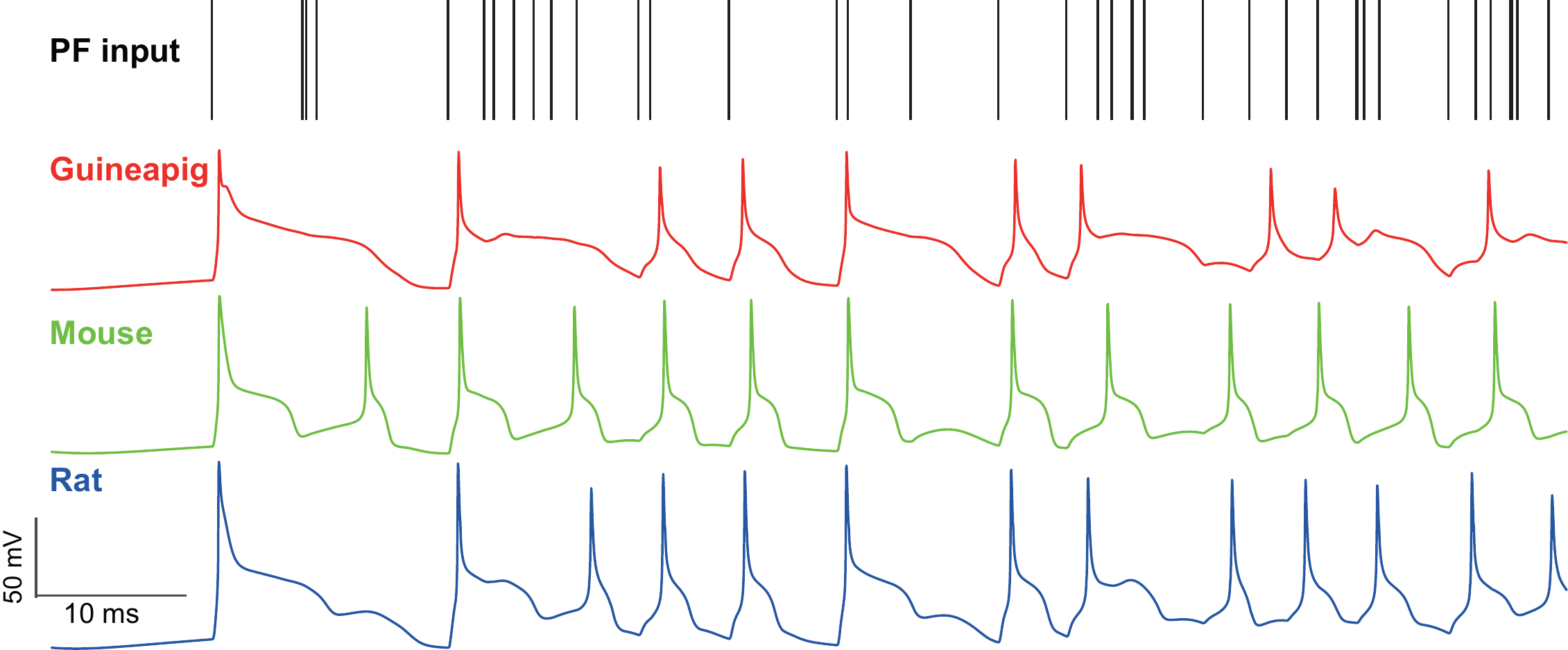}
	\caption{ Membrane potential traces recorded from soma of three example PCs of guinea-pig, mouse and rat in Branch model. Poisson stimulation at 500Hz.
    }
	\label{Membrane potential}
\end{figure*}

Furthermore, the performance of different reduced models to capture the temporal structure of spike trains shown in the full model is different. Under the same stimulation condition, the spike trains of mouse PC show more regularity in Horton and Shreve models, but rat show more regularity in Branch and Shreve models. However, guinea-pig only shows more regularity in Elect model under renewal process stimulation.  In addition, both full and reduce models have similar results under different stimulations protocols (Supplementary Fig.~2).

\begin{figure*}[tbp]
		\includegraphics[width=\columnwidth]{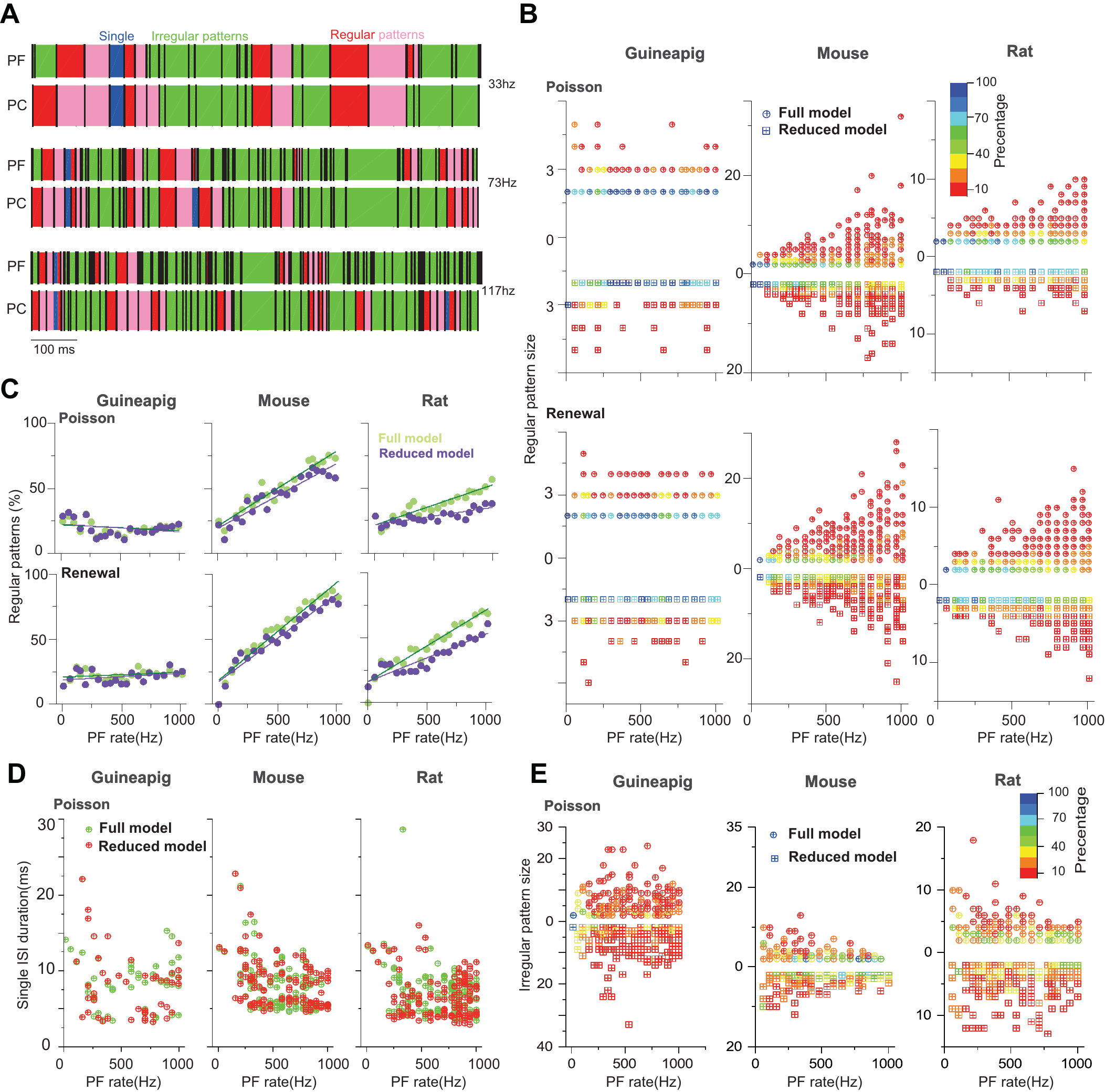}
		\centering
	\caption{ PC spiking pattern statistics with Branch method.  
    (A) Illustration of regular and irregular spiking patterns from spike trains. Each black bar indicates a spike. The start of regular patterns is colored in red, pink indicates successive ISIs in regular patterns, an irregular pattern is in green, and single ISI is in blue. Mouse PC is used.
    (B) Statistics of regular pattern size across a range of Poisson and renewal process stimulation for guinea-pig, mouse, and rat in both full and reduced models. Percentage of different size indicated by different colors, as there are more patterns in higher frequency.
    C) Percentage of regular patterns at different stimulation frequencies. 
  (D) Single ISI duration across a range of Poisson stimulation.
    (E) Statistics of irregular pattern size across a range of Poisson stimulation. Percentage of different size indicated by different colors.
    }
		\label{regular}
\end{figure*}

Cerebellar PCs are observed to generate regular spike trains in vivo~\cite{Hong2016Multiplexed, H1997Tonic}. PC simple spike trains contain highly regular spiking patterns and may transfer information coded by regular spike patterns to downstream deep cerebellar nuclei neurons~\cite{Shin2007Regular}. 

To see this effect, we applied a wide range of stimulation frequencies to extract the regular and irregular spiking patterns. A zoom-in illustration of firing activity is shown in Fig.~\ref{Membrane potential} under 500 Hz Poisson stimulation, which shows that guinea-pig has most irregular refractory periods as demonstrated in Fig.~\ref{ten-morphology} where there is a much slower hyperpolarization phase for guinea-pig, compared to mouse and rat under the same stimulus condition.

To characterize this in detail, we applied a threshold of 0.2 on the measured $\rm COV_2$ values similar to the previous study~\cite{Shin2007Regular} to extract a series of segments of regular spiking patterns in individual spike trains obtained by Branch model as illustrated in Fig.~\ref{regular} (A). 

Not surprisingly, when stimulus frequency is larger, regular patterns have small ISIs due to increased noise (Supplementary Fig.~4). However, the statistics of regular spiking patterns across species are quite different as shown in Fig.~\ref{regular} (C). Increasing of input frequency results in more regular patterns for mouse and rat, but there is no significant change for guinea-pig. 
Results of all four reduced models are similar to the full model for guinea-pig. For mouse, the difference is significantly larger in Horton and Shreve models, whereas this difference is very weak in Branch and Elect models. Rat PC shows more regular patterns in the full model than Branch and Shreve reduced models, but this difference is not obvious in Horton model (see Supplementary Fig.~7).

The size of the regular pattern is defined as the number of ISIs in regular patterns. Most of the patterns have only 2-3 ISIs for guinea-pig. Interestingly, for mouse and rat the regular pattern sizes are widely distributed at high frequencies, where there are quite a few longer regular patterns in mouse as in Fig.~\ref{regular} (B) and Supplementary Fig.~8.

Fig.~\ref{regular} (D) shows that guinea-pig has fewer single ISI and the single ISI duration distribution is more discrete at low frequencies. In contrary to regular pattern, irregular pattern sizes are widely distributed at low frequencies for mouse and rat, however, guinea-pig has larger irregular size as in Fig.~\ref{regular} (E) and Supplementary Fig.~9.

Thus, we conclude that reduced models can capture the PC temporal coding in a reasonable range, which is in particular important for maintaining precise timing patterns converted by PCs to theirs downstream deep cerebellar nuclei neurons to control motor patterns.

\subsection{ PC coding of modulated inputs}

In the previous sections, we demonstrated input-output relationships of PC firing activity by Poisson and renewal process stimulations with a constant firing rate. Recently, \textit{in vivo} studies show that the modulations of PC simple spike firing are related to a number of functions in terms of behavior, prediction, and sensory feedback~\cite{Streng2018Modulation,Cao2012Behavior}. In order to elucidate simple spike firing modulation, we simulated the dynamics of the PC in response to PF inputs with modulated firing amplitude and frequency. As the activities of PFs from granular cells are represented by modulations of stimuli, one can simply model PF inputs as sinusoidal modulations~\cite{Zampini2016Mechanisms} that can generate PF spike trains from a modulated renewal process.  By varying the amplitude and frequency of sinusoidal input, we investigated the ability of firing modulation of PC simple spikes under different reduced models.

\begin{figure*}[tbp]	
		\centering
		\includegraphics[width=\columnwidth]{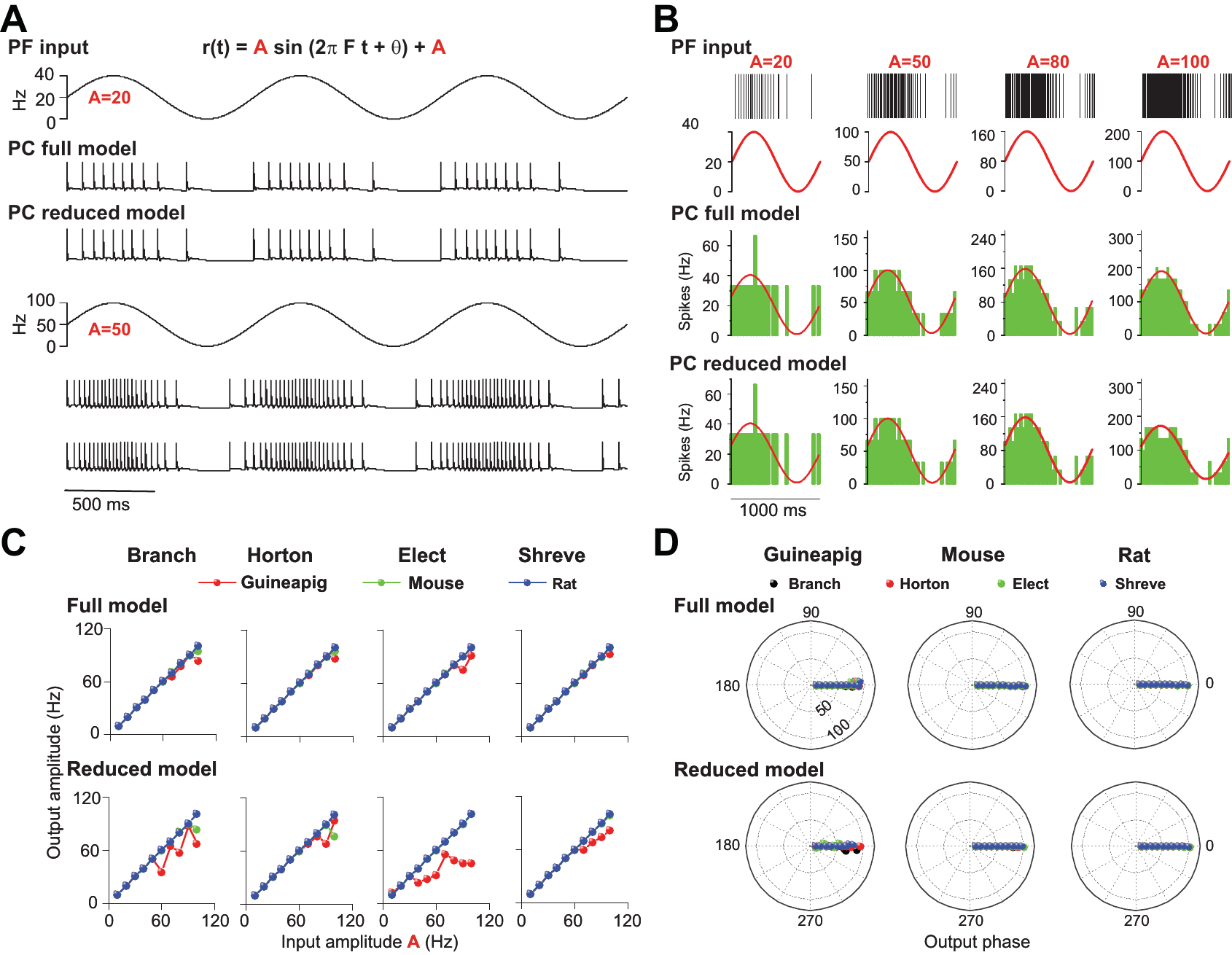}
	\caption{PC firing modulated by sinusoidal PF inputs with different amplitudes. 
    (A) Illustration of PF inputs and PC outputs. PF sinusoidal input at 1 Hz frequency with amplitude 20 (top) and 50 (bottom).  Modulated PC firing voltage traces over three cycles of input for full model and Branch model. 
    (B) Similar to (A) with four different amplitudes of 20, 50, 80, and 100 Hz inputs. (top) PF input spike trains and modulated firing rate. PC spike responses from (middle) full model and (bottom) Branch model. Mouse PC used in (A-B). 
    (C) Comparison of modulation amplitudes of PF input vs PC output in full and reduced models of four reduced schemes for guinea-pig (red), mouse (green), and rat (blue), respectively.
    (D) Similar to (C), but for phase change of PC firing modulation for full model and reduced models. 
    Sinusoidal stimulation frequency is 1Hz in all cases.  
    }
	\label{amplitude}
\end{figure*}

We first set out to analysis PC firing modulation by changing PF input amplitude, where frequency and phase of sinusoidal input is 1 Hz and 0 respectively.
Fig.~\ref{amplitude} (A) shows example results of mouse PC firing in response to modulated PF inputs at two different amplitudes, together with responses from reduced Branch model. The amplitude of PC modulation is expected to be equal to that of the oscillating input when the reduced model is accurate enough, which are shown in Fig.~\ref{amplitude} (B) with four tightly matched example responses for mouse PC. 

For the full model, the modulated amplitudes of mouse and rat PC responses are well matched to those of PF inputs (Fig.~\ref{amplitude} (C)). However, guinea-pig PC is different: well-matching is observed for lower amplitudes up to 90 Hz (80Hz in Elect method), but smaller than PF input amplitude for larger input amplitudes. Similarly, the PC firing of guinea-pig shows in phase with low amplitude PF inputs, but slightly leads phase at high input amplitude (large than 100Hz). However, mouse and rat PC modulation phases are in phase with PF inputs for all input frequencies (Fig.~\ref{amplitude} (D)). 

To investigate the firing modulation when PC morphological structures are changed, we simulated four corresponding reduced models. Comparing to the full model, guinea-pig shows lower modulation amplitudes for all reduced models. Mouse shows lower modulation amplitudes only in Branch and Horton reduced models with higher sinusoidal amplitude (90Hz). However, the modulation amplitudes of the rat are always matched to the PF input amplitudes in four reduced models (Fig.~\ref{amplitude} (C)). Moreover, in four reduced models, the modulation phases of mouse and rat are always in phase with PF inputs. However, guinea-pig shows leading phases at low input amplitudes and lagged phases at high input amplitudes. Therefore when the modulation amplitude changes, there is a large influence on PC modulated amplitudes and little influence on PC modulated phase. In particular, for guinea-pig,  the complex tree structure plays a key role in the firing modulation.

\begin{figure*}[thbp]	
	\centering
    \includegraphics[width=\columnwidth]{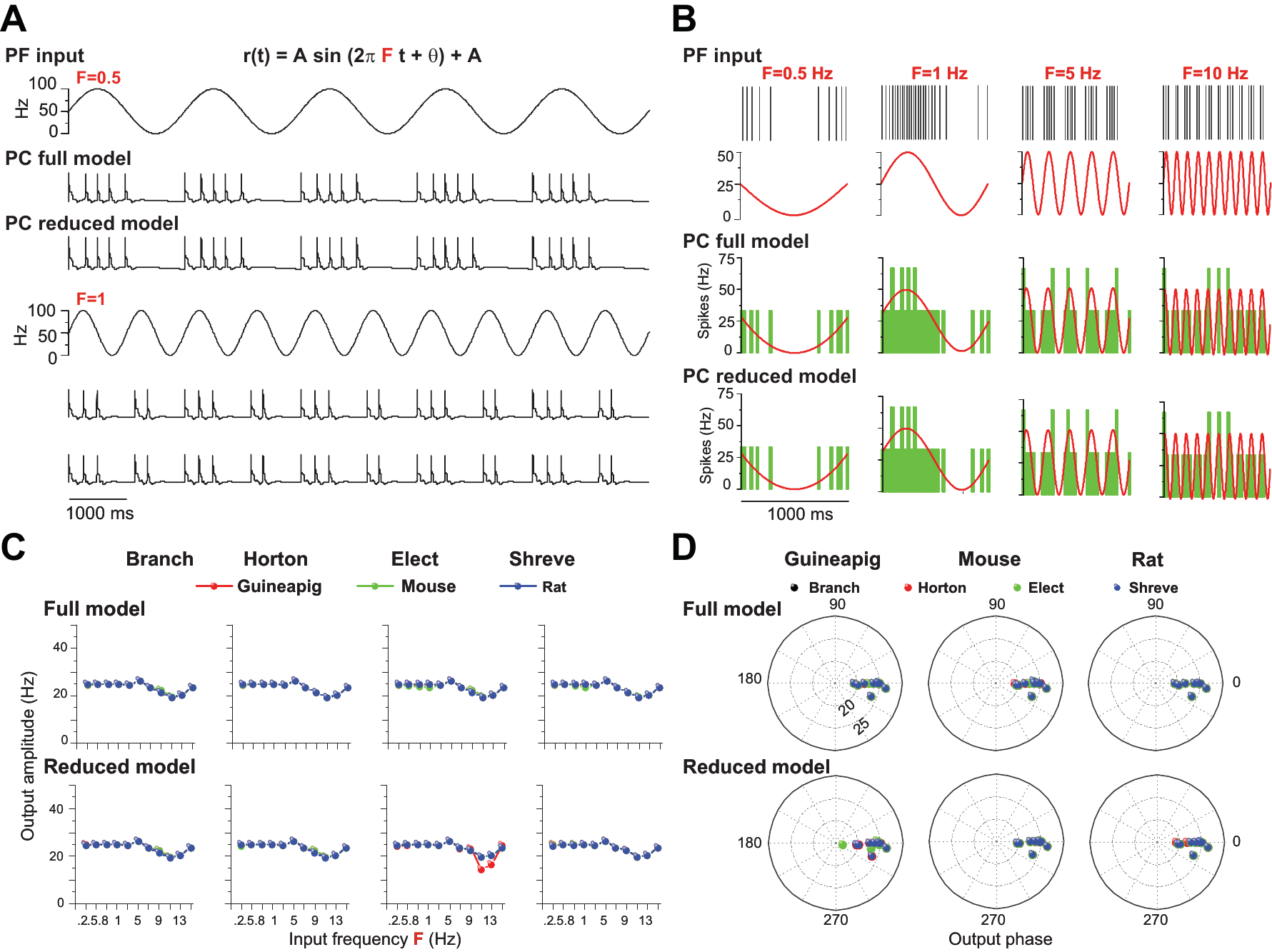}
	\caption{PC firing modulated by sinusoidal PF inputs with different frequencies. 
    (A) Illustration of PF inputs and PC outputs. PF sinusoidal input with amplitude 50 at frequencies of 0.5 Hz (top) and 1 (bottom).  Modulated PC firing voltage traces over five cycles of input for full model and Branch model. 
    (B) Similar to (A) with four different frequencies of 0.5, 1, 5, and 10 Hz inputs. (top) PF input spike trains and modulated firing rate. PC spike responses from (middle) full model and (bottom) Branch model. Mouse PC used in (A-B). 
    (C) Modulation amplitudes of PC output in full and reduced models of four reduced schemes for guinea-pig (red), mouse (green), and rat (blue), respectively, at different PF input frequencies.
    (D) Similar to (C), but for phase change of PC firing modulation for full model and reduced models. PF sinusoidal stimulation amplitude is 25Hz in (B-D). }
	\label{frequency}
\end{figure*}

Recently, theoretical studies showed the rate of PC tonic firing could be modulated by somatic injection of sinusoidal currents up to remarkably high frequency (1 kHz)~\cite{Ostojic2015Neuronal}. Here we replaced the current input with the modulated synaptic input as above, changed the frequency of PF sinusoidal input, and analyzed its effect on PC firing modulation as shown in Fig.~\ref{frequency}. 

Fig.~\ref{frequency} (A) shows example results of mouse PC firing modulation with given PF sinusoidal inputs with a fixed amplitude and phase as 0 for both full and Branch reduced models. PC firing is changing dynamically with different frequency changes. Averaged PC spiking responses over many cycles of inputs show that PC firing is well modulated and fitted with a wide range of sinusoidal input frequencies for both full and Branch reduced models (Fig.~\ref{frequency} (B)) for the amplitude of 25 Hz. 

PC output amplitudes can be characterized by fitted sinusoidal functions, which shows that the PC output amplitude of this modulation is equal to the PC input amplitude up to 3Hz for three typical PCs in the full models (Fig.~\ref{frequency} (C)). Then there is a small fluctuation of amplitude changes. Overall, these results are very captured by reduced models under different schemes. The only exception is for guinea-pig shown in Elect method (Fig.~\ref{frequency} (C)). In addition, the modulation phase changes are in phase with PF inputs at most of the frequencies (Fig.~\ref{frequency} (D)) for both full and reduced models across guinea-pig, mouse, and rat.

However, when the PF input amplitude is 50Hz, the PC output amplitude can only be modulated equally at sinusoidal low frequencies below 3Hz and decays for larger frequencies, but the amplitude of guinea-pig decays faster than mouse and rat (Supplementary Fig. 12).
Moreover,  for Elect reduced model, the PC output amplitude is much less than the input amplitude at all frequencies (Supplementary Fig. 12). 

Therefore, the PC firing could be modulated by simplifying the morphological structure at low amplitude when the sinusoidal frequency changes. 
However, the firing rate of guinea-pig PC could not be modulated at high amplitude in Elect reduced model when the frequency is changing.
We conclude that the Elect model, which has the highest degree of simplification, makes it difficult to be modulated at high amplitudes. Together with the previous results, we believe that the complex morphological structure contributes to PC firing modulation with given sinusoidal PF inputs.

\subsection{ Inhibition effect on PC coding }

Purkinje cells are the sole output of the cerebellar cortex. So far, we only consider the effects of excitatory inputs on PCs coding ability, since them receive the only excitatory signals of parallel fibers from granule cells which are the only output of granular layer. However, PCs also receive direct inhibitory input from several classes of molecular layer interneurons~\cite{brown2019molecular,barmack2008functions, he2015interneuron}, which contribute the firing activities of PCs significantly as well~\cite{brown2019molecular}.

\begin{figure*}[tbp]	
		\centering
		\includegraphics[width=\columnwidth]{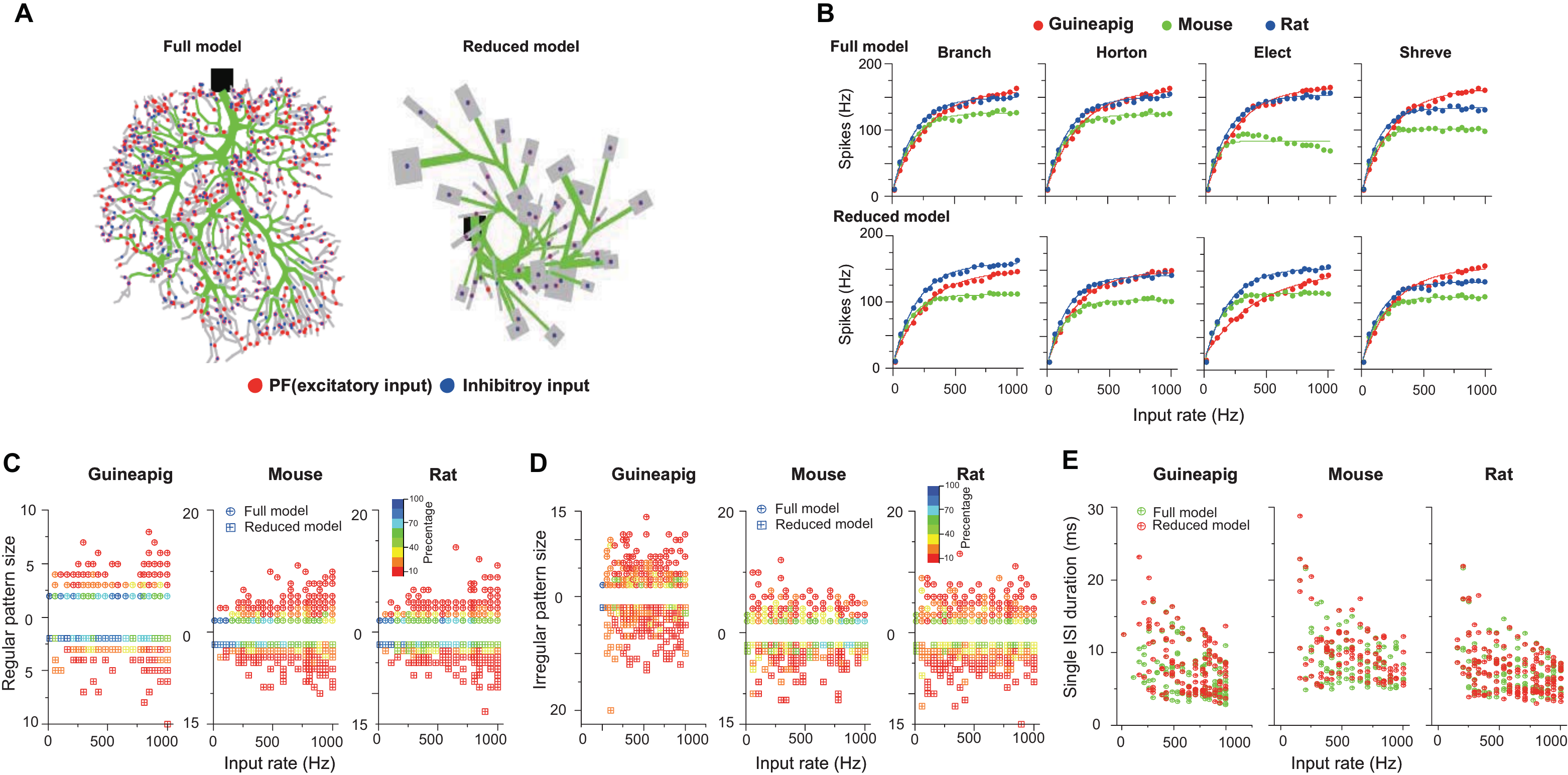}
	\caption{ PC firing activities with 500 inhibitory and 1000 excitatory inputs in Poisson stimulation. 
     (A) Full (left) and Branch (right) models receive excitatory (red) and inhibitory (blue) synapses. 
     (B) Comparison of firing response curves of three example PCs from guinea-pig, mouse and rate under four reduction schemes, Branch, Horton, Elect and Shreve, respectively. 
     (C) Statistics of regular pattern size across a range of Poisson stimulation for guinea-pig, mouse, and rat in full and Branch models. Percentage of different size is indicated by different colors. 
     (D) Similar as (C), but for irregular patterns.  
     (E) Single ISI duration across a range of Poisson stimulation.
     }
	\label{Inhibition}
\end{figure*}

Therefore, the effect of direct inhibition on PC firing activities were also studied by adding a population of inhibitory synaptic connections (See Methods). Indeed, we found there is a significant change of PC firing activities due to the inhibition input as shown in Fig.~\ref{Inhibition} with the same three example PCs of guinea-pig, mouse and rat under four reduction scheme under Poisson stimulation. Compared to firing activities in Fig.~\ref{tenpoissonfre} and spike timing patterns in Fig.~\ref{regular} where there is only PF input, there is a systematic influence from direct inhibitory input ( See Supplementary Fig. 10 and Supplementary Fig. 11). 
Although the results with inhibition are comparable to those without inhibition for most of cells, inhibitory input do play a functional role for adjusting firing activities, similar to the experimental observations~\cite{brown2019molecular}. The detailed performance with inhibitory inputs was computed in a similar way shown in Supplemental Tab. 1 for accuracy, Supplemental Tab. 2 for spike amplitude, and Supplemental Tab. 3 for spike width. Again, depending on the specific cells and species, there is a large diversity among different reduced methods.

\section{Summary and Discussion}

In this study, we investigated the coding capacity of Purkinje cells to excitatory parallel fiber input with different morphological reduction schemes. We proposed four reduction schemes to reduce the PC complex dendritic tree to a few components, and tested these reduced models with 10 specific detailed PCs from three species of guinea-pig, mouse and rat respectively. 

We showed by performance evaluation and simulation that reduced methods can balance accuracy and computational efficiency in different ways. 
We found that the Branch method has a better accuracy in most PCs, which is likely related to preserved volume in the reduced method. We also found that there is no direct relationship between accuracy and simplification. In Branch reduced model, PCs have the lowest level of simplification but only guinea-pig2 cell has the highest accuracy. In Elect reduced model, mouse1 cells show the highest accuracy and simplification at the same time. Rat1 cell has the highest degree of simplification but presents the lowest accuracy. In addition, the computing efficiency is proportional to the degree of the simplification. Therefore, the diversity of different reduction methods for performance implies that one has to choose a proper method depending on the questions to be addressed.

Compared to the Marosco model~\cite{Marasco2013Using}, we found Horton and Shreve model has a better accuracy in most PCs, in particular, Branch model is more accuracy than the Marosco model across all PCs. Furthermore, Elect model has a larger degree of simplification than Marosco model in most PCs.  We also found that the accuracy of the spike shape for four reduced models is higher than Marosco model in most PCs. However, there is no single method can achieve both good accuracy and simplification at the same time. These suggest that there is a trade-off of morphology reduction between accurate firing activity and efficient runtime of stimulation, which may need more systematically investigations at the level of neuronal network. 

Precise firing coding is a key property of Purkinje cells in the cerebellum that is mainly used to control the high-precise motor patterns with millisecond timescale~\cite{Ostojic2015Neuronal,Amir2011Cerebellum}. We found that mouse and rat PCs can response to generate higher firing rate than guinea-pig PCs with the same stimulation. This may imply rats and mice are able to react on millisecond timescale better than guinea-pig, and may have better performance for precise temporal control of motor-related tasks and conditioned behaviors. It is worth noting that the animal species are not uniquely represented by encoding of firing rate since the same species have different rate coding due to their differences in morphology. However, the detailed mechanisms of why different species show different dynamic behaviors during reduction remain unclear. More evidence is need to confirm and extend the current conclusion in the future. 

Besides firing rate coding, timing coding can be represented in PCs spiking patterns. With respect to simple spikes of PCs in response to PF excitatory inputs, PCs generate more regular spike trains than the stimulation sequences of Poisson process. For stimulation of renewal process, PF input sequences are more regular than PC spike trains at low frequencies. In contrary, PC spike trains are more regular than stimulation sequences at high frequencies.  Furthermore, rat and mouse PC spikes are more regular than guinea-pig at high frequencies. 
It is worth noting that increasing the rate of input results in a regularity in mouse and rat, but has no effect on guinea-pig. Inter-spike intervals have a linear relationship with refractory periods following each of spikes ~\cite{Guan2006The}. We found that guinea pig has most irregular refractory periods under the same stimulus condition (Fig.~\ref{Membrane potential}). Thus, we suggest that regularity is presumably determined by the refractory period.
Moreover, mouse and rat have similar result of the proportion of regular patterns in reduced models, but mouse is more regular. 
This phenomenon is likely to be determined by the refractory period.


We stress the importance of morphology because the morphology is known to define the feature of neuronal types and has significant influences on neuronal computation. A wide range of patterns of firing activity, dendritic processing and synaptic integration produced by differences in morphology determine the response of neurons to synaptic inputs~\cite{Mainen1996Influence,Einevoll2013Modelling, cannon2010stochastic}. 
Indeed, we found differences in neuronal morphology determine the response of PCs to synaptic inputs with a variation of firing rate, firing timing, and firing patterns. Here we mainly consider the case of PF synaptic inputs, since adding inhibitory inputs does not change the current results. However, it should be noted that the dynamics of neurons in the cerebellum is driven by a large diversity of synapses~\cite{Zampini2016Mechanisms}, and network dynamics can be reshaped by the various types of synaptic plasticities~\cite{Liu2009Embedding, Liu2011Learning}, future efforts are needed to study the effect of morphology on neuronal and network dynamics via different synaptic dynamics~\cite{Ostojic2015Neuronal, Yuste2001Morphological, Hering2001Dendritic}. 

\begin{figure*}[tbp]	
		\centering
		\includegraphics[width=\columnwidth]{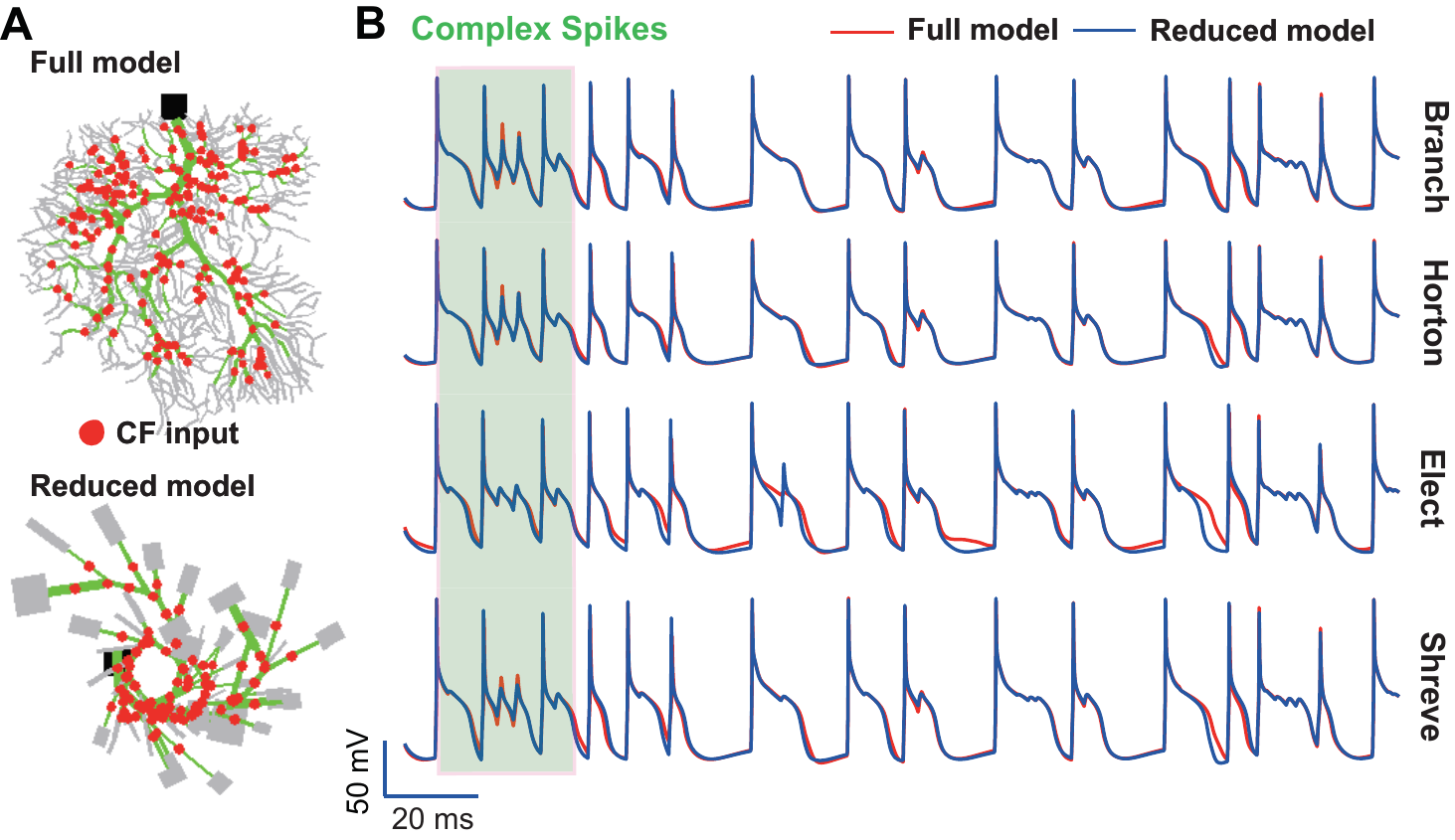}
	\caption{Complex spikes reproduced by reduced models. (A) Full and reduced morphology of guinea-pig receiving CF (red dot) synapses. (B) Membrane potential traces recorded from soma of guinea-pig PC in full (red) and four reduced models (blue). The shaded part indicates complex spikes. Poisson stimulation at 200Hz.
    }
	\label{CS}
\end{figure*}

It is well known that Purkinje cells affect motor behavior via both simple and complex spikes~\cite{Harmon2017Distinct,Khaliq2005Axonal}.  We mainly addressed simple spikes in this study and future work is needed to systematically analyze complex spikes discharged by PC. Simple spikes occur spontaneously and are modulated by synaptic inputs from granule cells. Complex spikes arise from climbing fibers and can induce plasticity of other afferents to Purkinje cells, therefore it plays the functional role of teaching or error signals during motor learning~\cite{Shogo2015Climbing, titley2017toward}. The granule cell inputs modulate simple spike firing up to 200 Hz, while climbing fibers trigger complex spikes at a remarkably low frequency (1 Hz)~\cite{Warnaar2015Duration}. In addition, complex spikes trigger prominent temporal pauses in the simple spike trains~\cite{Tang2017Heterogeneity}. While parallel fibers make synaptic contacts on spines in the spiny dendrite region, climbing fiber makes synaptic contact on the main and smooth dendrite regions~\cite{De1999Using,Achard2008Calcium}. We mainly simplified the spiny dendrites of the reduced models and preserved most of the morphological structure of smooth and main dendrites. While parallel fibers make synaptic contacts on spines in the spiny dendrite region, climbing fiber makes synaptic contact on the main and smooth dendrite regions. However, we found membrane potential traces can be reproduced well in four reduced models at climbing fiber input (Fig.~\ref{CS}). Thus, we speculate our reduced model can reproduce the properties of complex spikes very well. However, further systematic study is need to investigate complex spikes.

Therefore, reduced method can be redesigned to take into account this feature of the regional difference. It is possible that spiny dendrites have no effect on evoked complex spikes, so they can be eliminated when simplifying the morphological structure. 

Purkinje cell is a very unique cell type in that it has a very large, perhaps the largest and most dense neuronal morphology. However, there are many other cell types defined in other areas of the brain, in which neuronal morphologies are very different as well. Therefore, it is possible that one has to design specific reduction schemes according to the uniqueness of neuronal morphology for each cell type. For instance, the typical morphology of layer V neurons in the cortex can generate specific dendritic spikes that are, in particular, important to be kept when their morphologies are reduced~\cite{amsalem2018efficient}. Thus, one may not expect that there is a unique method for morphological reduction of all cell types in the neuronal system.



%

\section*{Acknowledgments}
We thank S. Jia for comments and suggestions. 
This work is supported by the National Key Research and Development Program of China (Grant No. 2016YFB0800601), the National Natural Science Foundation of China (Grant Nos. 61572385 and 6161101371), the Fundamental Research Funds for the Central Universities (Grant No. JB170309), the Science and Technology Projects of Xi'an, China (Grant No. 201809170CX11JC12), the International Talent Exchange Program of Beijing Municipal Commission of Science and Technology (Grant No. Z181100001018026), and the Royal Society Newton Advanced Fellowship (Grant No. NAF-R1-191082).



%
%
%

\bibliographystyle{IEEEtran}
\bibliography{pccoding}

\begin{thebibliography}{10}
\providecommand{\url}[1]{#1}
\csname url@samestyle\endcsname
\providecommand{\newblock}{\relax}
\providecommand{\bibinfo}[2]{#2}
\providecommand{\BIBentrySTDinterwordspacing}{\spaceskip=0pt\relax}
\providecommand{\BIBentryALTinterwordstretchfactor}{4}
\providecommand{\BIBentryALTinterwordspacing}{\spaceskip=\fontdimen2\font plus
\BIBentryALTinterwordstretchfactor\fontdimen3\font minus
  \fontdimen4\font\relax}
\providecommand{\BIBforeignlanguage}[2]{{%
\expandafter\ifx\csname l@#1\endcsname\relax
\typeout{** WARNING: IEEEtran.bst: No hyphenation pattern has been}%
\typeout{** loaded for the language `#1'. Using the pattern for}%
\typeout{** the default language instead.}%
\else
\language=\csname l@#1\endcsname
\fi
#2}}
\providecommand{\BIBdecl}{\relax}
\BIBdecl

\bibitem{Lapicque1907}
L.~Lapicque, ``Recherches quantitatives sur lexcitation \'electrique des nerfs
  trait\'eecomme une polarisation,'' \emph{J Physiol Pathol Gen(Paris)},
  vol.~9, pp. 620--635, 1907.

\bibitem{Burkitt2006A}
A.~N. Burkitt, ``A review of the integrate-and-fire neuron model: {I}.
  homogeneous synaptic input,'' \emph{Biological Cybernetics}, vol.~95, no.~1,
  pp. 1--19, 2006.

\bibitem{Burkitt2006A1}
------, ``A review of the integrate-and-fire neuron model: {II}. inhomogeneous
  synaptic input and network properties.'' \emph{Biological Cybernetics},
  vol.~95, no.~2, pp. 97--112, 2006.

\bibitem{Mainen1996Influence}
Z.~F. Mainen and T.~J. Sejnowski, ``Influence of dendritic structure on firing
  pattern in model neocortical neurons,'' \emph{Nature}, vol. 382, no. 6589,
  pp. 363--366, 1996.

\bibitem{Koch1999Biophysics}
C.~Koch, \emph{\rm{Biophysics of Computation: Information Processing in Single
  Neurons}}.\hskip 1em plus 0.5em minus 0.4em\relax Oxford University Press,
  1999.

\bibitem{Poirazi2003Arithmetic}
P.~Poirazi, T.~Brannon, Mel, and W.~Bartlett, ``Arithmetic of subthreshold
  synaptic summation in a model {CA1} pyramidal cell,'' \emph{Neuron}, vol.~37,
  no.~6, pp. 977--987, 2003.

\bibitem{Ostojic2015Neuronal}
S.~Ostojic, G.~Szapiro, E.~Schwartz, B.~Barbour, N.~Brunel, and V.~Hakim,
  ``Neuronal morphology generates high-frequency firing resonance.''
  \emph{Journal of Neuroscience}, vol.~35, no.~18, p. 7056, 2015.

\bibitem{amsalem2018efficient}
O.~Amsalem, G.~Eyal, N.~Rogozinski, and I.~Segev, ``An efficient analytical
  reduction of nonlinear detailed neuron models,'' \emph{bioRxiv}, 2018.

\bibitem{Herz2006Modeling}
A.~V. Herz, T.~Gollisch, C.~K. Machens, and D.~Jaeger, ``Modeling single-neuron
  dynamics and computations: a balance of detail and abstraction,''
  \emph{Science}, vol. 314, no. 5796, pp. 80--5, 2006.

\bibitem{Rall1959Branching}
W.~Rall, ``Branching dendritic trees and motoneuron membrane resistivity,''
  \emph{Experimental Neurology}, vol.~1, no.~5, pp. 491--527, 1959.

\bibitem{Hodgkin1990A}
A.~L. Hodgkin and A.~F. Huxley, ``A quantitative description of membrane
  current and its application to conduction and excitation in nerve,''
  \emph{Bulletin of Mathematical Biology}, vol.~52, no. 1-2, pp. 25--71, 1990.

\bibitem{Brown2011Virtual}
S.~A. Brown, I.~I. Moraru, J.~C. Schaff, and L.~M. Loew, ``Virtual neuron: a
  strategy for merged biochemical and electrophysiological modeling,''
  \emph{Journal of Computational Neuroscience}, vol.~31, no.~2, pp. 385--400,
  2011.

\bibitem{Armin2012Automated}
B.~Armin, M.~B. Stemmler, A.~V.~M. Herz, and R.~Arnd, ``Automated optimization
  of a reduced layer 5 pyramidal cell model based on experimental data,''
  \emph{Journal of Neuroscience Methods}, vol. 210, no.~1, pp. 22--34, 2012.

\bibitem{Marasco2013Using}
A.~Marasco, A.~Limongiello, and M.~Migliore, ``Using strahler's analysis to
  reduce up to 200-fold the run time of realistic neuron models,''
  \emph{Scientific Reports}, vol.~3, no.~1, p. 2934, 2013.

\bibitem{Ito1984The}
M.~Ito, ``The modifiable neuronal network of the cerebellum,'' \emph{Japanese
  Journal of Physiology}, vol.~34, no.~5, p. 781, 1984.

\bibitem{Zeeuw1997Volume}
C.~I.~D. Zeeuw, P.~Strata, and J.~Voogd, ``Volume 114. the cerebellum: From
  structure to control,'' 1997.

\bibitem{Amir2011Cerebellum}
K.~Amir and D.~S. Zee, ``Cerebellum and ocular motor control,'' \emph{Frontiers
  in Neurology}, vol.~2, no.~53, p.~53, 2011.

\bibitem{Manto2012Consensus}
M.~Manto, A.~B. Conforto, G.~J.~M. Delgado, M.~Gerwig, C.~Habas, N.~Hagura, and
  R.~B. Ivry, ``Consensus paper: Roles of the cerebellum in motor control—the
  diversity of ideas on cerebellar involvement in movement,''
  \emph{Cerebellum}, vol.~11, no.~2, pp. 457--487, 2012.

\bibitem{Du1995Learning}
L.~S. Du, J.~L. Raymond, T.~J. Sejnowski, and S.~G. Lisberger, ``Learning and
  memory in the vestibulo-ocular reflex.'' \emph{Annual Review Neuroscience},
  vol.~18, no.~1, p. 409, 1995.

\bibitem{Hirata2012Direct}
Y.~Hirata, K.~Katagiri, and Y.~Tanaka, ``Direct causality between
  single-purkinje cell activities and motor learning revealed by a
  cerebellum-machine interface utilizing vor adaptation paradigm,''
  \emph{Cerebellum}, vol.~11, no.~2, pp. 455--456, 2012.

\bibitem{Blazquez2004The}
P.~M. Blazquez, Y.~Hirata, and S.~M. Highstein, ``The vestibulo-ocular reflex
  as a model system for motor learning: what is the role of the cerebellum?''
  \emph{Cerebellum}, vol.~3, no.~3, pp. 188--192, 2004.

\bibitem{Koekkoek2003Cerebellar}
S.~K.~E. Koekkoek, H.~C. Hulscher, B.~R. Dortland, R.~A. Hensbroek,
  Y.~Elgersma, T.~J.~H. Ruigrok, and C.~I.~D. Zeeuw, ``Cerebellar {LTD} and
  learning-dependent timing of conditioned eyelid responses,'' \emph{Science},
  vol. 301, no. 5640, pp. 1736--9, 2003.

\bibitem{Jim2004Role}
L.~Jim\'enezdíaz, J.~D. Navarrol\'opez, A.~Gruart, and J.~M. Delgadogarcía,
  ``Role of cerebellar interpositus nucleus in the genesis and control of
  reflex and conditioned eyelid responses.'' \emph{Journal of Neuroscience},
  vol.~24, no.~41, p. 9138, 2004.

\bibitem{Wolf2009Evaluating}
U.~Wolf, M.~J. Rapoport, and T.~A. Schweizer, ``Evaluating the affective
  component of the cerebellar cognitive affective syndrome.'' \emph{Journal of
  Neuropsychiatry \& Clinical Neurosciences}, vol.~21, no.~3, p. 245, 2009.

\bibitem{Wagner2017Cerebellar}
M.~J. Wagner, T.~H. Kim, J.~Savall, M.~J. Schnitzer, and L.~Luo, ``Cerebellar
  granule cells encode the expectation of reward,'' \emph{Nature}, vol. 544,
  no. 7648, pp. 96--100, 2017.

\bibitem{Ito2008Control}
M.~Ito, ``Control of mental activities by internal models in the cerebellum,''
  \emph{Nature Reviews Neuroscience}, vol.~9, no.~4, pp. 304--313, 2008.

\bibitem{Strick2009Cerebellum}
P.~L. Strick, R.~P. Dum, and J.~A. Fiez, ``Cerebellum and nonmotor function.''
  \emph{Annual Review of Neuroscience}, vol.~32, no.~32, pp. 413--434, 2009.

\bibitem{Tsai2012Autistic}
P.~T. Tsai, H.~Court, C.~Yunxiang, G.~C. Emily, A.~R. Sadowski, J.~M. Leech,
  S.~Jason, J.~N. Crawley, W.~G. Regehr, and S.~Mustafa, ``Autistic-like
  behaviour and cerebellar dysfunction in purkinje cell {Tsc1} mutant mice,''
  \emph{Nature}, vol. 488, no. 7413, pp. 647--51, 2012.

\bibitem{Bostan2018The}
A.~C. Bostan and P.~L. Strick, ``The basal ganglia and the cerebellum: nodes in
  an integrated network,'' \emph{Journal of the American Chemical Society},
  vol.~9, no.~1, pp. 1--11, 2018.

\bibitem{Raymond2018Computational}
J.~L. Raymond and J.~F. Medina, ``Computational principles of supervised
  learning in the cerebellum,'' \emph{Annual Review of Neuroscience}, vol.~41,
  no.~1, p. 233, 2018.

\bibitem{Loewenstein2005Bistability}
Y.~Loewenstein, S.~Mahon, P.~Chadderton, K.~Kitamura, H.~Sompolinsky, Y.~Yarom,
  and M.~Häusser, ``Bistability of cerebellar purkinje cells modulated by
  sensory stimulation.'' \emph{Nature Neuroscience}, vol.~8, no.~2, pp.
  202--11, 2005.

\bibitem{Popa2016The}
L.~S. Popa, M.~L. Streng, A.~L. Hewitt, and T.~J. Ebner, ``The errors of our
  ways: Understanding error representations in cerebellar-dependent motor
  learning,'' \emph{Cerebellum}, vol.~15, no.~2, p.~93, 2016.

\bibitem{Robinson2001The}
F.~R. Robinson and A.~F. Fuchs, ``The role of the cerebellum in voluntary eye
  movements,'' \emph{Annual Review of Neuroscience}, vol.~24, no.~1, pp.
  981--1004, 2001.

\bibitem{Chen2016The}
S.~Chen, G.~J. Augustine, and P.~Chadderton, ``The cerebellum linearly encodes
  whisker position during voluntary movement,'' \emph{Elife}, vol.~5, p.
  e10509, 2016.

\bibitem{Hewitt2015Changes}
A.~L. Hewitt, L.~S. Popa, and T.~J. Ebner, ``Changes in purkinje cell simple
  spike encoding of reach kinematics during adaption to a mechanical
  perturbation,'' \emph{Journal of Neuroscience}, vol.~35, no.~3, pp.
  1106--1124, 2015.

\bibitem{Popa2015Predictive}
L.~S. Popa, A.~L. Hewitt, and T.~J. Ebner, ``Predictive and feedback
  performance errors are signaled in the simple spike discharge of individual
  purkinje cells.'' \emph{Journal of Neuroscience}, vol.~32, no.~44, p. 15345,
  2015.

\bibitem{Streng2018Modulation}
M.~L. Streng, L.~S. Popa, and T.~J. Ebner, ``Modulation of sensory prediction
  error in purkinje cells during visual feedback manipulations.'' \emph{Nature
  Communications}, vol.~9, no.~1, 2018.

\bibitem{Horton1945EROSIONAL}
R.~E. Horton, ``Erosional development of streams and their drainage basins;
  hydrophysical approach to quantitative morphology,'' \emph{Journal of the
  Japanese Forestry Society}, vol.~56, no.~3, pp. 275--370, 1945.

\bibitem{Shreve1967Infinite}
R.~L. Shreve, ``Infinite topologically random channel networks,'' \emph{Journal
  of Geology}, vol.~75, no.~2, pp. 178--186, 1967.

\bibitem{Rapp1994Physiology}
M.~Rapp, I.~Segev, and Y.~Yarom, ``Physiology, morphology and detailed passive
  models of guinea-pig cerebellar purkinje cells,'' \emph{J Physiol}, vol. 474,
  no.~1, pp. 101--118, 1994.

\bibitem{De1994An}
E.~D. Schutter and J.~M. Bower, ``An active membrane model of the cerebellar
  purkinje cell. {I}. simulation of current clamps in slice.'' \emph{Journal of
  Neurophysiology}, vol.~71, no.~1, pp. 375--400, 1994.

\bibitem{Miyasho2001Low}
T.~Miyasho, H.~Takagi, H.~Suzuki, S.~Watanabe, M.~Inoue, Y.~Kudo, and
  H.~Miyakawa, ``Low-threshold potassium channels and a low-threshold calcium
  channel regulate $\rm{Ca^{2+}}$ spike firing in the dendrites of cerebellar
  purkinje neurons: a modeling study.'' \emph{Brain Research}, vol. 891, no.
  1–2, pp. 106--115, 2001.

\bibitem{Gao2012Distributed}
Z.~Gao, B.~J. van Beugen, and C.~I. De~Zeeuw, ``Distributed synergistic
  plasticity and cerebellar learning.'' \emph{Nature Reviews Neuroscience},
  vol.~13, no.~9, p. 619, 2012.

\bibitem{Masoli2017Synaptic}
S.~Masoli and E.~D’Angelo, ``Synaptic activation of a detailed purkinje cell
  model predicts voltage-dependent control of burst-pause responses in active
  dendrites,'' \emph{Frontiers in Cellular Neuroscience}, vol.~11, p. 278,
  2017.

\bibitem{De1994An1}
E.~D. Schutter and J.~M. Bower, ``An active membrane model of the cerebellar
  purkinje cell. {II}. simulation of synaptic responses.'' \emph{Journal of
  Neurophysiology}, vol.~71, no.~1, p. 401, 1994.

\bibitem{he2015interneuron}
Q.~He, I.~Duguid, B.~Clark, P.~Panzanelli, B.~Patel, P.~Thomas, J.~M. Fritschy,
  and T.~G. Smart, ``Interneuron-and gaba a receptor-specific inhibitory
  synaptic plasticity in cerebellar purkinje cells,'' \emph{Nature
  communications}, vol.~6, p. 7364, 2015.

\bibitem{Napper1988Number}
R.~M.~A. Napper and R.~J. Harvey, ``Number of parallel fiber synapses on an
  individual purkinje cell in the cerebellum of the rat,'' \emph{Journal of
  Comparative Neurology}, vol. 274, no.~2, pp. 168--177, 1988.

\bibitem{Hoxha2016Modulation}
E.~Hoxha, F.~Tempia, P.~Lippiello, and M.~C. Miniaci, ``Modulation, plasticity
  and pathophysiology of the parallel fiber-purkinje cell synapse,''
  \emph{Frontiers in Synaptic Neuroscience}, vol.~8, 2016.

\bibitem{Jaeger2003No}
D.~Jaeger, ``No parallel fiber volleys in the cerebellar cortex: evidence from
  cross-correlation analysis between purkinje cells in a computer model and in
  recordings from anesthetized rats,'' \emph{Journal of Computational
  Neuroscience}, vol.~14, no.~3, pp. 311--327, 2003.

\bibitem{Su2012Target}
J.~Su, R.~S. Stenbjorn, K.~Gorse, K.~Su, K.~F. Hauser, S.~Ricardblum,
  T.~Pihlajaniemi, and M.~A. Fox, ``Target-derived matricryptins organize
  cerebellar synapse formation through $\rm{\alpha3\beta1}$ integrins,''
  \emph{Cell Reports}, vol.~2, no.~2, p. 223, 2012.

\bibitem{Brown2002The}
E.~N. Brown, R.~Barbieri, V.~Ventura, R.~E. Kass, and L.~M. Frank, ``The
  time-rescaling theorem and its application to neural spike train data
  analysis,'' \emph{Neural Computation}, vol.~14, no.~2, pp. 325--346, 2002.

\bibitem{Pillow2009Time}
J.~Pillow, ``Time-rescaling methods for the estimation and assessment of
  non-poisson neural encoding models,'' in \emph{23rd Annual Conference on
  Neural Information Processing Systems}, 2009, pp. 1473--1481.

\bibitem{Zampini2016Mechanisms}
V.~Zampini, J.~K. Liu, M.~A. Diana, P.~P. Maldonado, N.~Brunel, and
  S.~Dieudonn\'e, ``Mechanisms and functional roles of glutamatergic synapse
  diversity in a cerebellar circuit,'' \emph{Elife}, vol.~5, 2016.

\bibitem{Valera2012Adaptation}
A.~M. Valera, F.~Doussau, B.~Poulain, B.~Barbour, and P.~Isope, ``Adaptation of
  granule cell to purkinje cell synapses to high-frequency transmission,''
  \emph{Journal of Neuroscience}, vol.~32, no.~9, pp. 3267--80, 2012.

\bibitem{Jelitai2016Dendritic}
M.~Jelitai, P.~Puggioni, T.~Ishikawa, A.~Rinaldi, and I.~Duguid, ``Dendritic
  excitation–inhibition balance shapes cerebellar output during motor
  behaviour,'' \emph{Nature Communications}, vol.~7, p. 13722, 2016.

\bibitem{Bryant2010Cerebellar}
J.~L. Bryant, J.~D. Boughter, S.~M. Gong, and D.~H. Heck, ``Cerebellar cortical
  output encodes temporal aspects of rhythmic licking movements and is
  necessary for normal licking frequency.'' \emph{European Journal of
  Neuroscience}, vol.~32, no.~1, pp. 41--52, 2010.

\bibitem{Cao2017Cerebellar}
Y.~Cao, Y.~Liu, D.~Jaeger, and D.~H. Heck, ``Cerebellar purkinje cells generate
  highly correlated spontaneous slow-rate fluctuations,'' \emph{Front Neural
  Circuits}, vol.~11, p.~67, 2017.

\bibitem{Ivry2004The}
R.~B. Ivry and R.~M. Spencer, ``The neural representation of time,''
  \emph{Current Opinion in Neurobiology}, vol.~14, no.~2, p. 225, 2004.

\bibitem{shin2006dynamic}
S.~L. Shin and E.~D. Schutter, ``Dynamic synchronization of purkinje cell
  simple spikes,'' \emph{Journal of neurophysiology}, 2006.

\bibitem{Shin2007Regular}
S.~L. Shin, F.~E. Hoebeek, M.~Schonewille, C.~I.~D. Zeeuw, A.~Aertsen, and
  E.~D. Schutter, ``Regular patterns in cerebellar purkinje cell simple spike
  trains,'' \emph{Plos One}, vol.~2, no.~5, p. e485, 2007.

\bibitem{Hong2016Multiplexed}
S.~Hong, M.~Negrello, M.~Junker, A.~Smilgin, P.~Thier, and E.~D. Schutter,
  ``Multiplexed coding by cerebellar purkinje neurons,'' \emph{Elife}, vol.~5,
  no. 2016, 2016.

\bibitem{H1997Tonic}
M.~\'H\"ausser and B.~A. Clark, ``Tonic synaptic inhibition modulates neuronal
  output pattern and spatiotemporal synaptic integration,'' \emph{Neuron},
  vol.~19, no.~3, p. 665, 1997.

\bibitem{Cao2012Behavior}
Y.~Cao, S.~K. Maran, M.~Dhamala, D.~Jaeger, and D.~H. Heck, ``Behavior related
  pauses in simple spike activity of mouse purkinje cells are linked to spike
  rate modulation,'' \emph{Journal of Neuroscience}, vol.~32, no.~25, p. 8678,
  2012.

\bibitem{brown2019molecular}
A.~M. Brown, M.~Arancillo, T.~Lin, D.~R. Catt, J.~Zhou, E.~P. Lackey, T.~L.
  Stay, Z.~Zuo, J.~J. White, and R.~V. Sillitoe, ``Molecular layer interneurons
  shape the spike activity of cerebellar purkinje cells,'' \emph{Scientific
  Reports}, vol.~9, no.~1, p. 1742, 2019.

\bibitem{barmack2008functions}
N.~H. Barmack and V.~Yakhnitsa, ``Functions of interneurons in mouse
  cerebellum,'' \emph{Journal of Neuroscience}, vol.~28, no.~5, pp. 1140--1152,
  2008.

\bibitem{Guan2006The}
S.~Guan, S.~Ma, Y.~Zhu, and J.~Wang, ``The postnatal development of refractory
  periods and threshold potentials at cerebellar purkinje neurons,''
  \emph{Brain Research}, vol. 1097, no.~1, pp. 59--64, 2006.

\bibitem{Einevoll2013Modelling}
G.~T. Einevoll, C.~Kayser, N.~K. Logothetis, and S.~Panzeri, ``Modelling and
  analysis of local field potentials for studying the function of cortical
  circuits,'' \emph{Nature Reviews Neuroscience}, vol.~14, no.~11, pp.
  770--785, 2013.

\bibitem{cannon2010stochastic}
R.~C. Cannon, C.~O'Donnell, and M.~F. Nolan, ``Stochastic ion channel gating in
  dendritic neurons: morphology dependence and probabilistic synaptic
  activation of dendritic spikes,'' \emph{PLoS computational biology}, vol.~6,
  no.~8, p. e1000886, 2010.

\bibitem{Liu2009Embedding}
J.~K. Liu and D.~V. Buonomano, ``Embedding multiple trajectories in simulated
  recurrent neural networks in a self-organizing manner,'' \emph{Journal of
  Neuroscience}, vol.~29, no.~42, pp. 13\,172--13\,181, 2009.

\bibitem{Liu2011Learning}
J.~K. Liu, ``Learning rule of homeostatic synaptic scaling: Presynaptic
  dependent or not,'' \emph{Neural Computation}, vol.~23, no.~12, pp.
  3145--3161, 2011.

\bibitem{Yuste2001Morphological}
R.~Yuste and T.~Bonhoeffer, ``Morphological changes in dendritic spines
  associated with long-term synaptic plasticity,'' \emph{Annual Review of
  Neuroscience}, vol.~24, no.~1, pp. 1071--1089, 2001.

\bibitem{Hering2001Dendritic}
H.~Hering and M.~Sheng, ``Dendritic spines: structure, dynamics and
  regulation,'' \emph{Nature Reviews Neuroscience}, vol.~2, no.~12, pp. 880--8,
  2001.

\bibitem{Harmon2017Distinct}
T.~C. Harmon, U.~Magaram, D.~L. Mclean, and I.~M. Raman, ``Distinct responses
  of purkinje neurons and roles of simple spikes during associative motor
  learning in larval zebrafish,'' \emph{Elife}, vol.~6, 2017.

\bibitem{Khaliq2005Axonal}
Z.~M. Khaliq and I.~M. Raman, ``Axonal propagation of simple and complex spikes
  in cerebellar purkinje neurons,'' \emph{Journal of Neuroscience}, vol.~25,
  no.~2, p. 454, 2005.

\bibitem{Shogo2015Climbing}
O.~Shogo and J.~F. Medina, ``Climbing fibers encode a temporal-difference
  prediction error during cerebellar learning in mice:,'' \emph{Nature
  Neuroscience}, vol.~18, no.~12, pp. 1798--803, 2015.

\bibitem{titley2017toward}
H.~K. Titley, N.~Brunel, and C.~Hansel, ``Toward a neurocentric view of
  learning,'' \emph{Neuron}, vol.~95, no.~1, pp. 19--32, 2017.

\bibitem{Warnaar2015Duration}
P.~Warnaar, J.~Couto, M.~Negrello, M.~Junker, A.~Smilgin, A.~Ignashchenkova,
  M.~Giugliano, P.~Thier, and E.~D. Schutter, ``Duration of purkinje cell
  complex spikes increases with their firing frequency,'' \emph{Frontiers in
  Cellular Neuroscience}, vol.~9, p. 122, 2015.

\bibitem{Tang2017Heterogeneity}
T.~Tang, J.~Xiao, C.~Y. Suh, A.~Burroughs, N.~L. Cerminara, L.~Jia, S.~P.
  Marshall, A.~K. Wise, R.~Apps, and I.~Sugihara, ``Heterogeneity of purkinje
  cell simple spike-complex spike interactions: zebrin-\ and non-zebrin-\
  related variations,'' \emph{Journal of Physiology}, vol. 595, no.~15, 2017.

\bibitem{De1999Using}
E.~D. Schutter, ``Using realistic models to study synaptic integration in
  cerebellar purkinje cells,'' \emph{Rev Neurosci}, vol.~10, no. 3-4, pp.
  233--245, 1999.

\bibitem{Achard2008Calcium}
P.~Achard and E.~D. Schutter, ``Calcium, synaptic plasticity and intrinsic
  homeostasis in purkinje neuron models,'' \emph{Frontiers in Computational
  Neuroscience}, vol.~2, no.~2, p.~8, 2008.

\end{thebibliography}

\end{document}


%
\title{Coding Capacity of Purkinje Cells with Different Schemes of Morphological Reduction: \\
Supplemental Figures}

\author{\IEEEauthorblockN{Lingling~An\dag, Yuanhong~Tang\dag,  Quan~Wang\dag, Qingqi~Pei\dag, Ran~Wei\dag, Huiyuan~Duan\dag, Jian~K.~Liu\ddag \\}
	\IEEEauthorblockA{\dag  School of Computer Science and Technology, Xidian University, Xi'an, China\\
		\ddag Centre for Systems Neuroscience,  
        Department of Neuroscience, Psychology and Behaviour, University of Leicester, Leicester, UK\\
		Correspondence:: an.lingling@gmail.com, jian.liu@leicester.ac.uk
        }
}

\maketitle
\IEEEpeerreviewmaketitle

\begin{table}[bhpb]
	\renewcommand{\arraystretch}{1}
	\caption{ Accuracy of reduced models  with 500 inhibition input. Values are mean $\pm$ STD. STD is calculated from 21  sets of Poisson stimulation frequency from 10 to 1K Hz.} 
	\label{Table inhibition accuracy}
	\begin{tabular}{|p{0.9cm}<{\centering}|p{1.7cm}<{\centering}|p{1.7cm}<{\centering}|p{1.7cm}<{\centering}|p{1cm}<{\centering}|p{1cm}<{\centering}|p{1cm}<{\centering}|p{1cm}<{\centering}|p{0.9cm}<{\centering}|p{0.9cm}<{\centering}|p{0.9cm}<{\centering}|}
		\hline
        &\textbf{Guinea-pig1}& \textbf{Guinea-pig2} & \textbf{Guinea-pig3}&\textbf{Mouse1}& \textbf{Mouse2 }& \textbf{Mouse3 }& \textbf{Mouse4 }&\textbf{ rat1} &\textbf{  rat2} &\textbf{ rat3}\\
        \hline
       \textbf{Branch}&\boldmath{$0.957\pm{0.016}$}&$0.927\pm{0.032}$ &$0.962\pm{0.020}$ &$0.926\pm{0.045}$ &         $0.922\pm{0.042}$&\boldmath{$0.989\pm{0.008}$}&\boldmath{$0.980\pm{0.015}$ }&$0.970\pm{0.017}$&$0.935\pm{0.021}$&$0.953\pm{0.030}$\\
       \hline
         \textbf{Horton} & $0.949\pm{0.018}$&\boldmath{$0.929\pm{0.023}$ }&\boldmath{$0.971\pm{0.012}$}  &$0.880\pm{0.054}$&$0.914\pm{0.039}$ & $0.971\pm{0.021}$& $0.937\pm{0.027}$ & $0.967\pm{0.023}$&$0.946\pm{0.016}$ &\boldmath{$0.961\pm{0.021}$} \\
         \hline	
          \textbf{Elect}& $0.867\pm{0.034}$& $0.836\pm{0.050}$&$0.865\pm{0.041}$&$0.836\pm{0.087}$&$0.881\pm{0.058}$&$0.958\pm{0.032}$& $0.912\pm{0.026}$& \boldmath{$0.977\pm{0.024}$}&$0.744\pm{0.065}$&$0.915\pm{0.025}$\\
          	\hline	
            \textbf{Shreve}& $0.946\pm{0.019}$&$0.919\pm{0.030}$&$0.966\pm{0.013}$& \boldmath{$0.955\pm{0.026}$}&\boldmath{$0.932\pm{0.034}$}&$0.985\pm{0.014}$&$0.980\pm{0.012}$&$0.964\pm{0.018}$&\boldmath{ $0.977\pm{0.017}$}&$0.919\pm{0.041}$\\
		\hline
	\end{tabular}
\end{table}
\begin{table}[hpb]
	\renewcommand{\arraystretch}{1}
	\caption{The change of spike amplitude (mv) in reduced models with 500 inhibition inputs. Values are mean $\pm$ STD. STD is calculated from 21  sets of Poisson stimulation frequency from 10 to 1K Hz. }
	\label{Table inhibition amplitude}
	\begin{tabular}{|p{0.9cm}<{\centering}|p{1.7cm}<{\centering}|p{1.7cm}<{\centering}|p{1.7cm}<{\centering}|p{1cm}<{\centering}|p{1cm}<{\centering}|p{1cm}<{\centering}|p{1cm}<{\centering}|p{0.9cm}<{\centering}|p{0.9cm}<{\centering}|p{0.9cm}<{\centering}|}
		\hline
        &\textbf{Guinea-pig1}& \textbf{Guinea-pig2} & \textbf{Guinea-pig3}&\textbf{Mouse1}& \textbf{Mouse2 }& \textbf{Mouse3 }& \textbf{Mouse4 }&\textbf{ rat1} &\textbf{  rat2} &\textbf{ rat3}\\
        \hline
       \textbf{Branch}&\boldmath{$1.4\pm{0.9}$}&$2.7\pm{1.4}$ &$2.7\pm{1.3}$ &$0.4\pm{0.2}$ &         \boldmath{$3.4\pm{1.2}$}&\boldmath{$0.1\pm{0.2}$}&$0.2\pm{0.1}$ &$2.2\pm{0.9}$&$2.5\pm{0.9}$&$0.6\pm{0.3}$\\
       \hline
         \textbf{Horton} & $2.3\pm{0.9}$&$2.2\pm{0.6}$ &\boldmath{$1.5\pm{0.6}$}  &$0.5\pm{0.2}$&$3.6\pm{1.4}$ & $0.4\pm{0.2}$& \boldmath{$0.1\pm{0.1}$} & $0.8\pm{0.3}$&\boldmath{$0.3\pm{0.5}$} &\boldmath{$0.5\pm{0.2}$} \\
         \hline	
          \textbf{Elect}& $2.1\pm{2.1}$& \boldmath{$1.3\pm{0.7}$}& \boldmath{$1.4\pm{0.9}$}&$0.5\pm{0.1}$&$6.6\pm{2.1}$&$0.9\pm{0.6}$& $3.8\pm{0.1}$& \boldmath{$0.4\pm{0.6}$}&$1.3\pm{0.7}$&$2.1\pm{0.9}$\\
          	\hline	
            \textbf{Shreve}& $3.1\pm{1.1}$&$3.0\pm{1.1}$&$2.1\pm{0.9}$& \boldmath{$0.1\pm{0.1}$}&$3.4\pm{1.3}$&$0.3\pm{0.2}$&$0.6\pm{0.3}$&$1.2\pm{0.3}$& $0.3\pm{0.7}$&$1.3\pm{0.5}$\\
		\hline
	\end{tabular}
\end{table}

\begin{table}[tpb]
	\renewcommand{\arraystretch}{1}
	\caption{ The chance of spike width (ms) in reduced models with 500 inhibition inputs. Values are mean $\pm$ STD. STD is calculated from 21  sets of Poisson stimulation frequency from 10 to 1K Hz. }
	\label{Table inhibition width}
	\begin{tabular}{|p{0.9cm}<{\centering}|p{1.7cm}<{\centering}|p{1.7cm}<{\centering}|p{1.7cm}<{\centering}|p{1cm}<{\centering}|p{1cm}<{\centering}|p{1cm}<{\centering}|p{1cm}<{\centering}|p{0.9cm}<{\centering}|p{0.9cm}<{\centering}|p{0.9cm}<{\centering}|}
		\hline
        &\textbf{Guinea-pig1}& \textbf{Guinea-pig2} & \textbf{Guinea-pig3}&\textbf{Mouse1}& \textbf{Mouse2 }& \textbf{Mouse3 }& \textbf{Mouse4 }&\textbf{ rat1} &\textbf{  rat2} &\textbf{ rat3}\\
        \hline
       \textbf{Branch}&$0.23\pm{0.29}$&$0.52\pm{0.23}$ &$0.27\pm{0.18}$ &$0.20\pm{0.07}$ &       $1.45\pm{0.92}$&$0.18\pm{0.12}$&$0.3\pm{0.13}$ &$0.31\pm{0.49}$&$0.56\pm{0.25}$&\boldmath{$0.09\pm{0.1}$}\\
       \hline
         \textbf{Horton} & $0.29\pm{0.43}$&$0.69\pm{0.36}$ &\boldmath{$0.16\pm{0.23}$}  &$0.26\pm{0.07}$&$1.1\pm{0.93}$ & $0.1\pm{0.11}$& $0.7\pm{0.3}$ & \boldmath{$0.1\pm{0.1}$}&$0.36\pm{0.21}$&$0.1\pm{0.12}$\\
         \hline	
          \textbf{Elect}& $0.68\pm{0.52}$& $1.07\pm{1.02}$& $0.64\pm{0.68}$&\boldmath{$0.1\pm{0.12}$}&\boldmath{$0.6\pm{0.6}$}&$0.36\pm{0.48}$& $0.43\pm{0.49}$& $0.24\pm{0.37}$&$0.55\pm{0.32}$&$0.71\pm{0.9}$\\
          	\hline	
            \textbf{Shreve}& \boldmath{$0.21\pm{0.14}$}&\boldmath{$0.49\pm{0.34}$}&$0.19\pm{0.34}$& $0.19\pm{0.05}$&$0.97\pm{0.8}$&\boldmath{$0.1\pm{0.1}$}&\boldmath{$0.25\pm{0.2}$}&$0.44\pm{0.47}$& \boldmath{$0.26\pm{0.18}$}&$0.34\pm{0.39}$\\
		\hline
	\end{tabular}
\end{table}

\newpage

\begin{figure*}[thbp]
\centering
\includegraphics[width=\columnwidth]{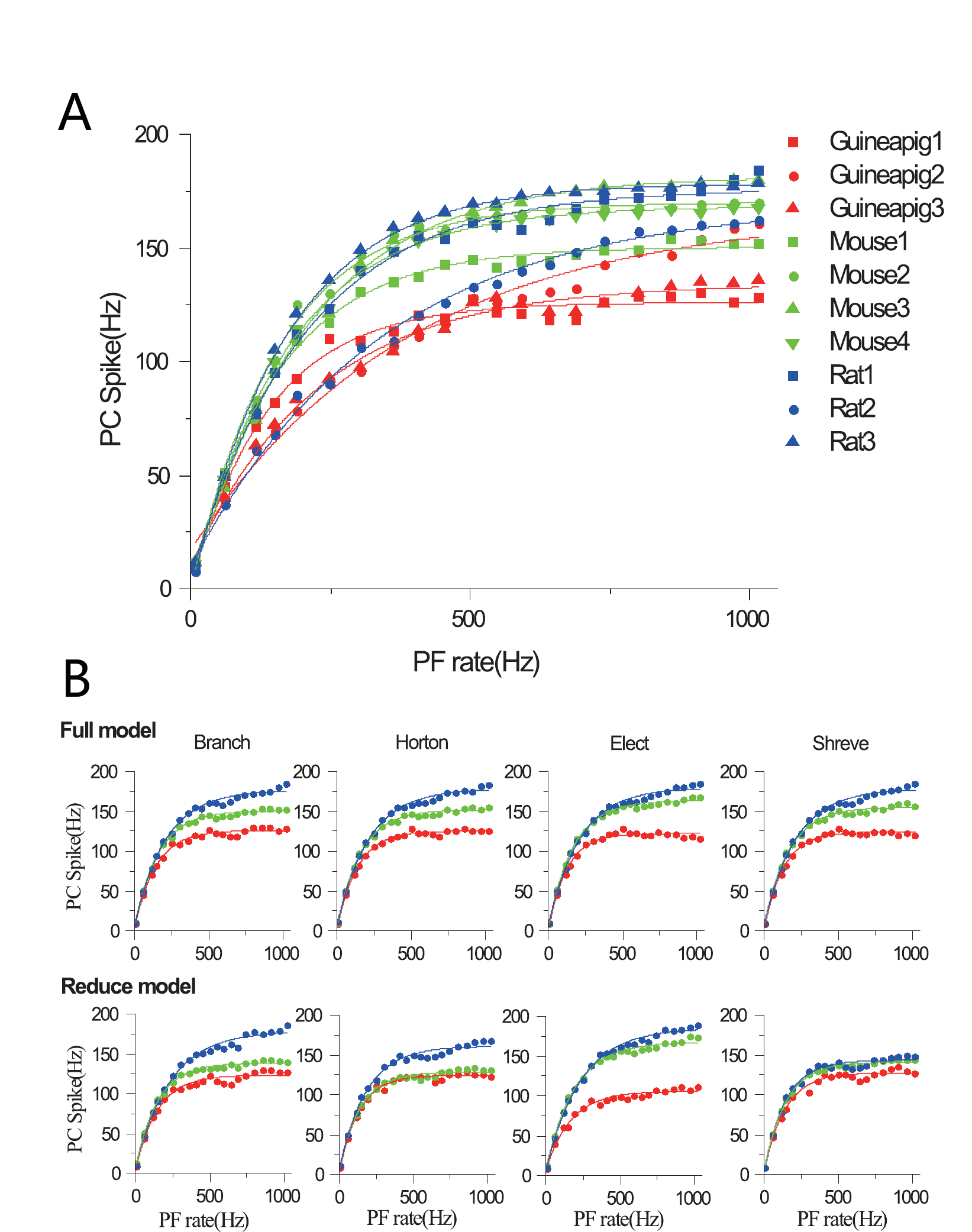}
    \centering	
    \label{tenrenewalfre}
\caption{ PC firing response in full and reduced model. (A)  Firing response curves of ten PCs with full morphology with renewal process stimulation frequencies 10Hz to 1000Hz. (B). Comparison of firing response curves of three example PCs from guinea-pig, mouse and rate under four reduction schemes, Branch, Horton, Elect and Shreve, respectively. }
\end{figure*}


\begin{figure*}[thbp]
\raggedright
\includegraphics[width=0.9\columnwidth]{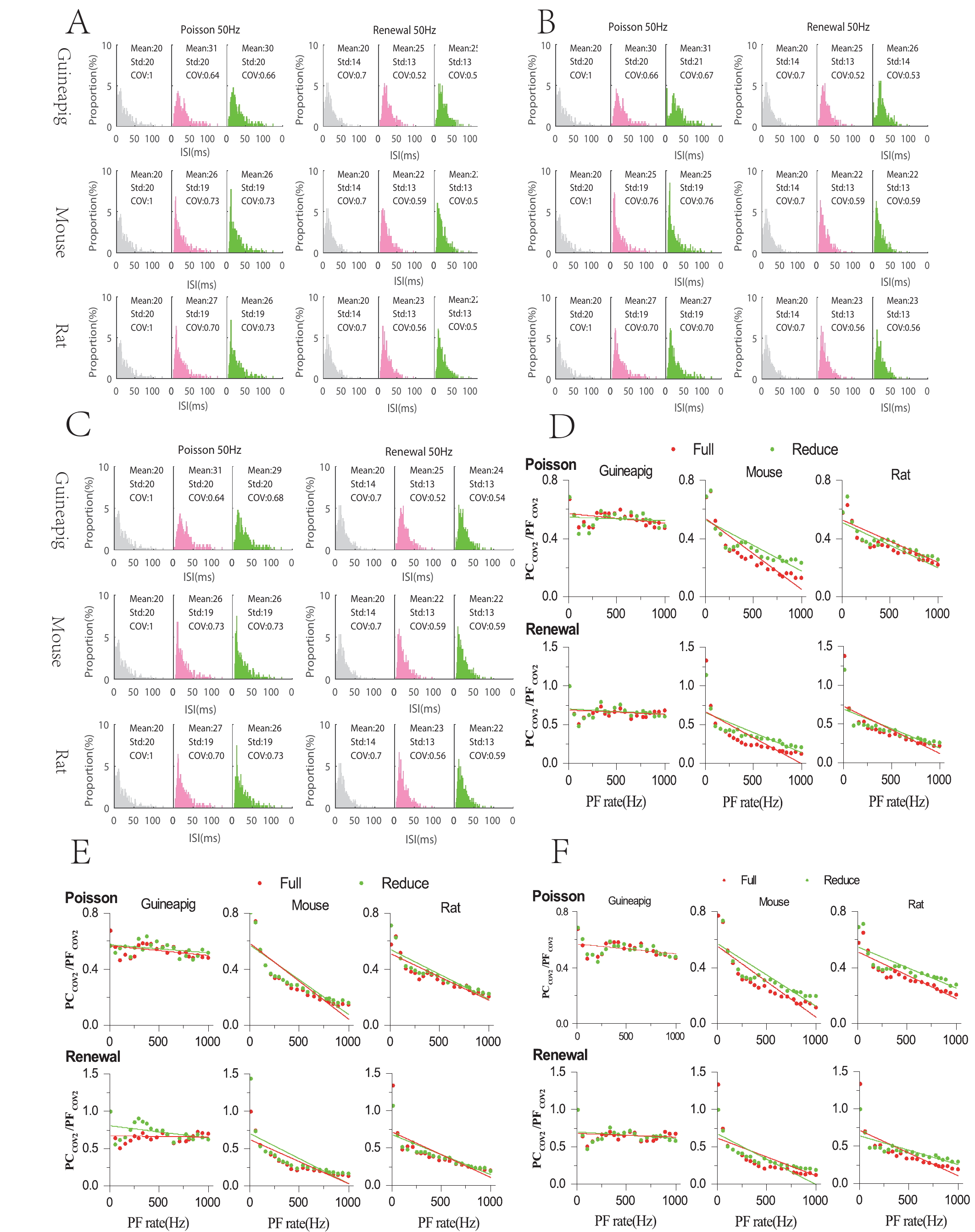}
    \centering	
\caption{PC temporal response under Poisson and renewal stimulation. (A), (B) and (C) are Horton, Elect and Shreve reduced model, respectively.  ISI distribution of spike trains from PF (gray), PC full model (light red) and PC reduced model (green) under Poisson and renewal process stimulation for guinea-pig (top), mouse (middle), rat (bottom). All stimulation frequencies are 50Hz for 10 seconds. P$> 0.1$, Wilcoxon Rank-sum test. (D), (E) and (F) are Horton, Elect and Shreve reduced model, respectively.  $\rm PC_{cov2}/PF_{cov2}$ showing the regularity between PF inputs and PC outputs for full (red) and reduced model (green) of guinea-pig, mouse and rat. Stimulations are Poisson and renewal process with different frequencies from 10 to 1K Hz. 
}
\end{figure*}

\begin{figure*}[thbp]
 \centering	
\includegraphics[width=\columnwidth]{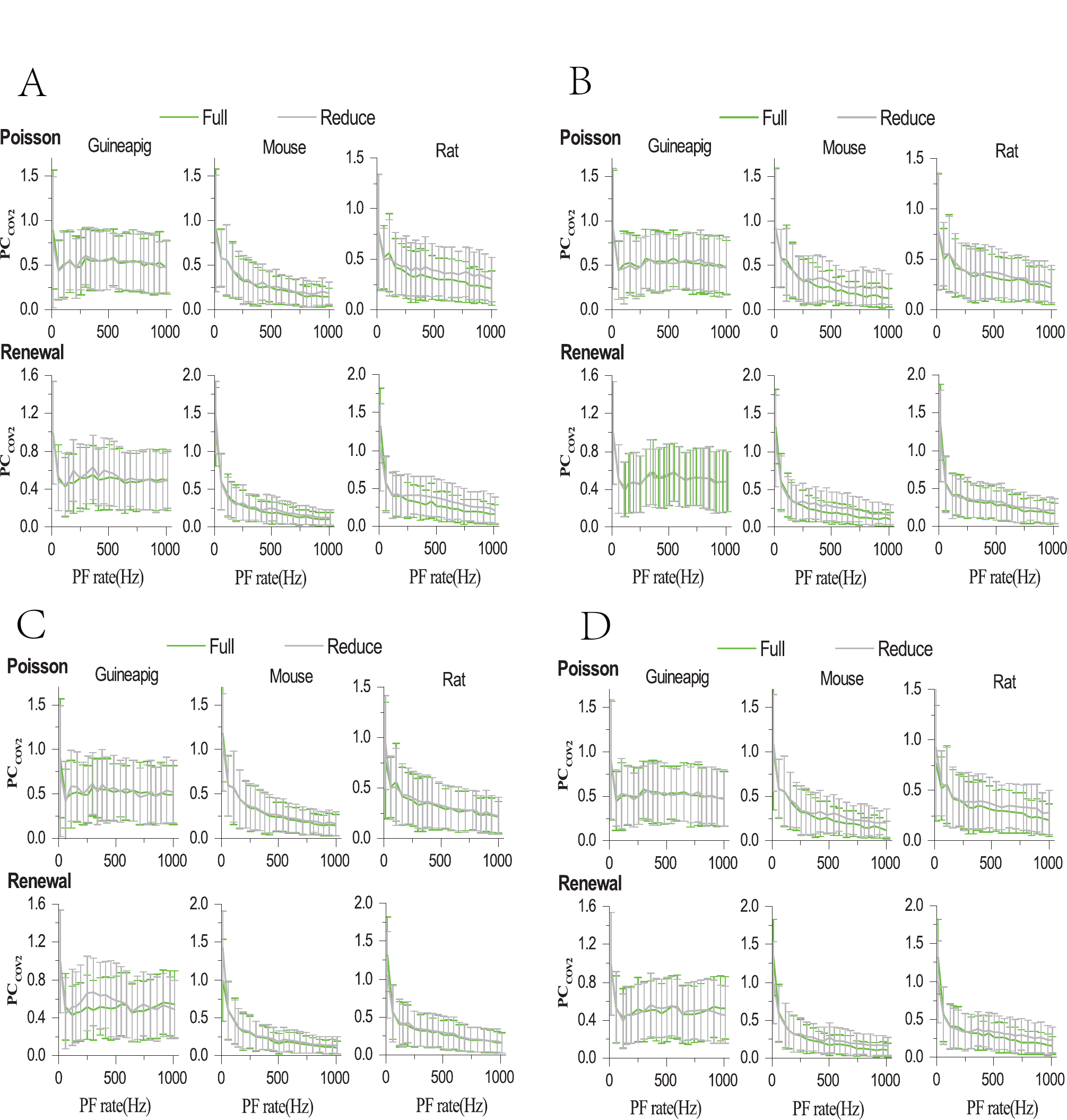}
    \centering	
    \label{}
\caption{(A)(Branch), (B)(Horton), (C)(Elect) and (D)(Shreve) show the $mean\pm std$  of $cov_2$ of the PC spike trains in full (green) and reduced (gray) model with Poisson and Renewal stimulation frequencies 10Hz to 1KHz.}
\end{figure*}

\begin{figure*}[thbp]
 \centering	
		\includegraphics[width=\columnwidth]{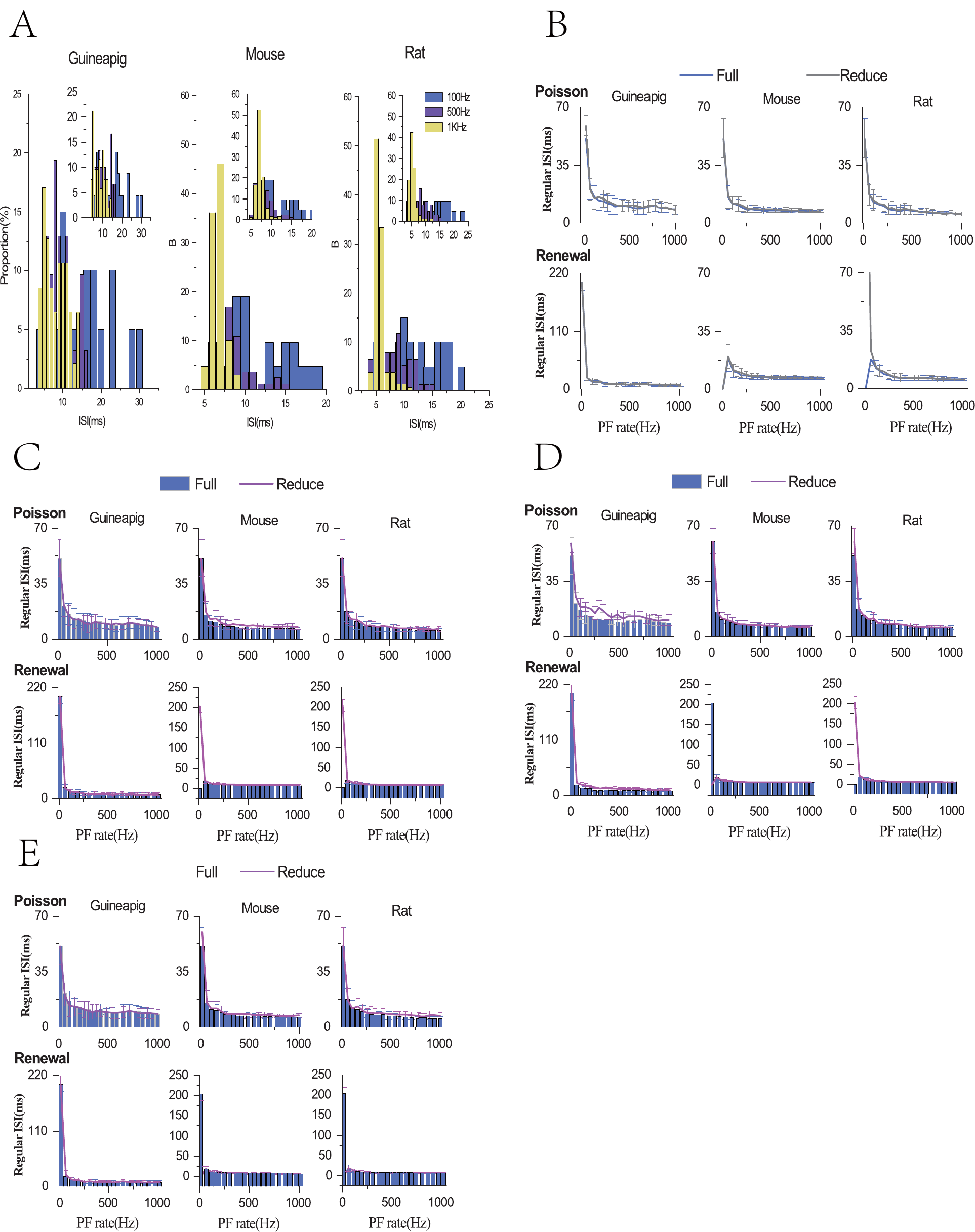}
		\centering
\caption{PC spiking patterns. 
(A). The mouse PC spiking regular pattern for 100Hz, 500Hz and 1KHz with Poisson stimulation in full and Branch reduced model(inset). (B)(Branch), (C)(Horton), (D)(Elect) and (E)(Shreve) show  $mean\pm std$  of regular patterns with Poisson and renewal stimulation in full and reduced modes.}
\end{figure*}

\begin{figure*}[thbp]
 \centering	
		
		\includegraphics[width=\columnwidth]{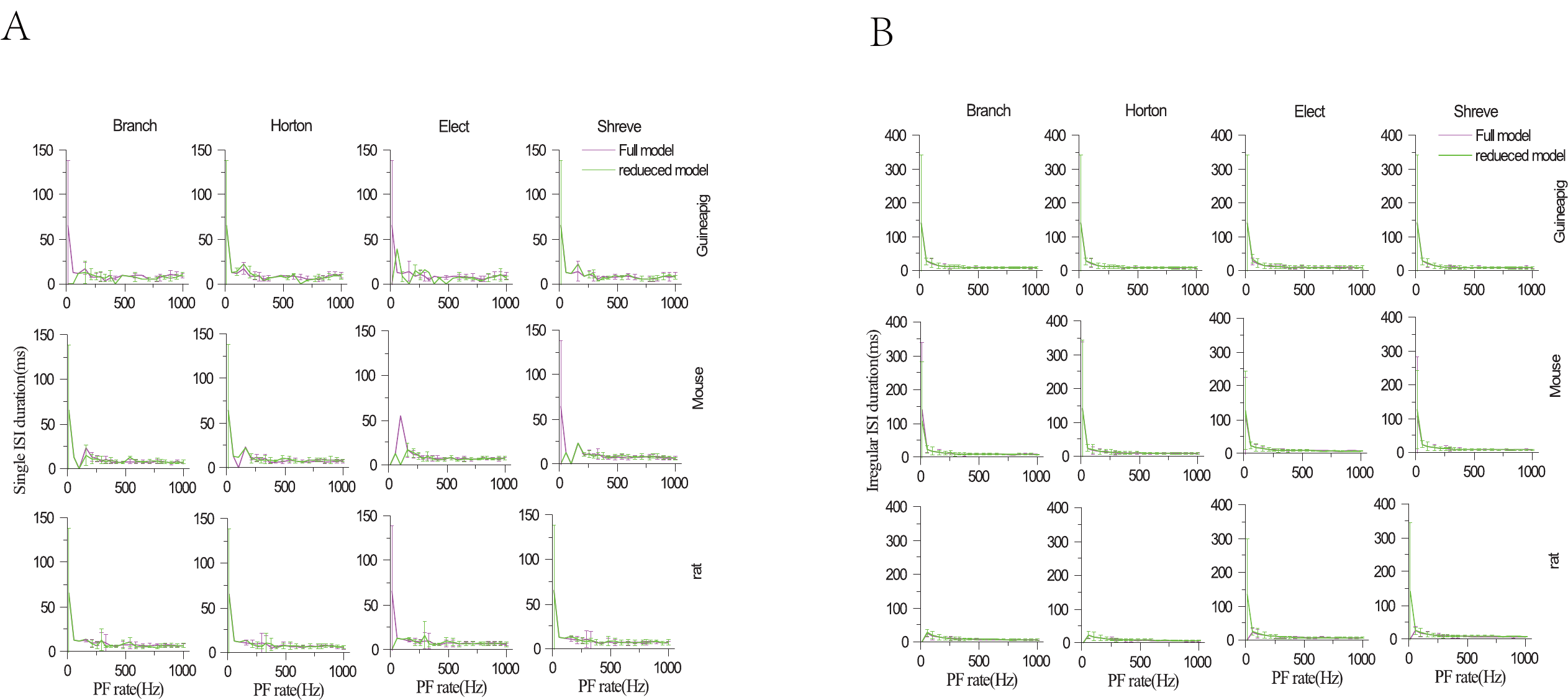}
		\centering
\caption{
PC spiking patterns. 
(A). Single ISI duration ($mean\pm std$) with Poisson stimulation for guinea pig (top), mouse (middle) and rat (bottom) in full and reduced models. (B) Irregular patterns.}
\end{figure*}

\begin{figure*}[thbp]
 \centering	
		
		\includegraphics[width=\columnwidth]{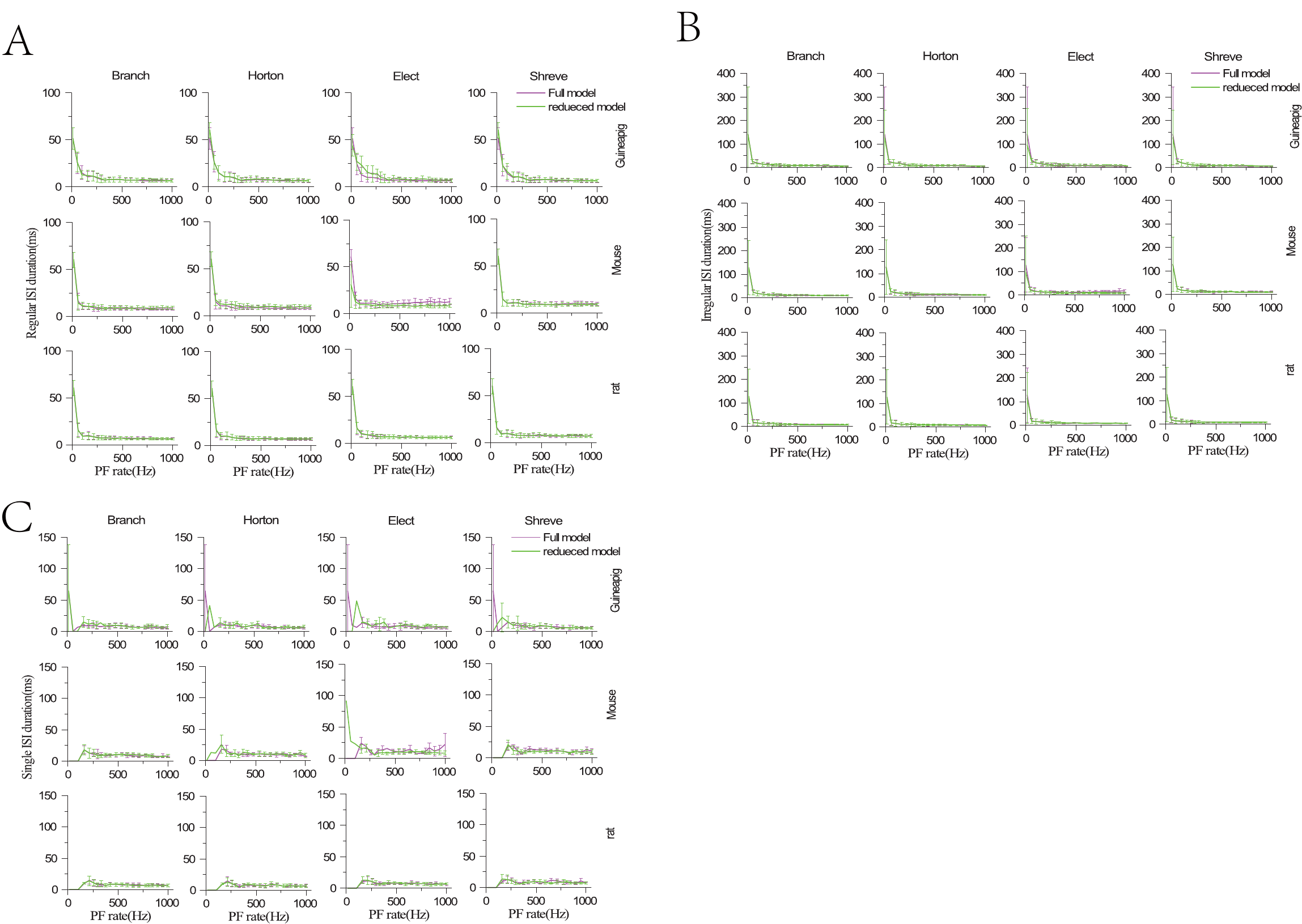}
		\centering
\caption{ PC spiking patterns with inhibition input. 
(A) Regular pattern ISI duration ( $mean\pm std$) with Poisson stimulation for guinea pig (top), mouse (middle) and rat (bottom) in full and reduced model. (B) Irregular pattern ISI duration distribution. (C) Single ISI duration distribution.
}
\end{figure*}

\begin{figure*}[thbp]
    \centering	
\includegraphics[width=\columnwidth]{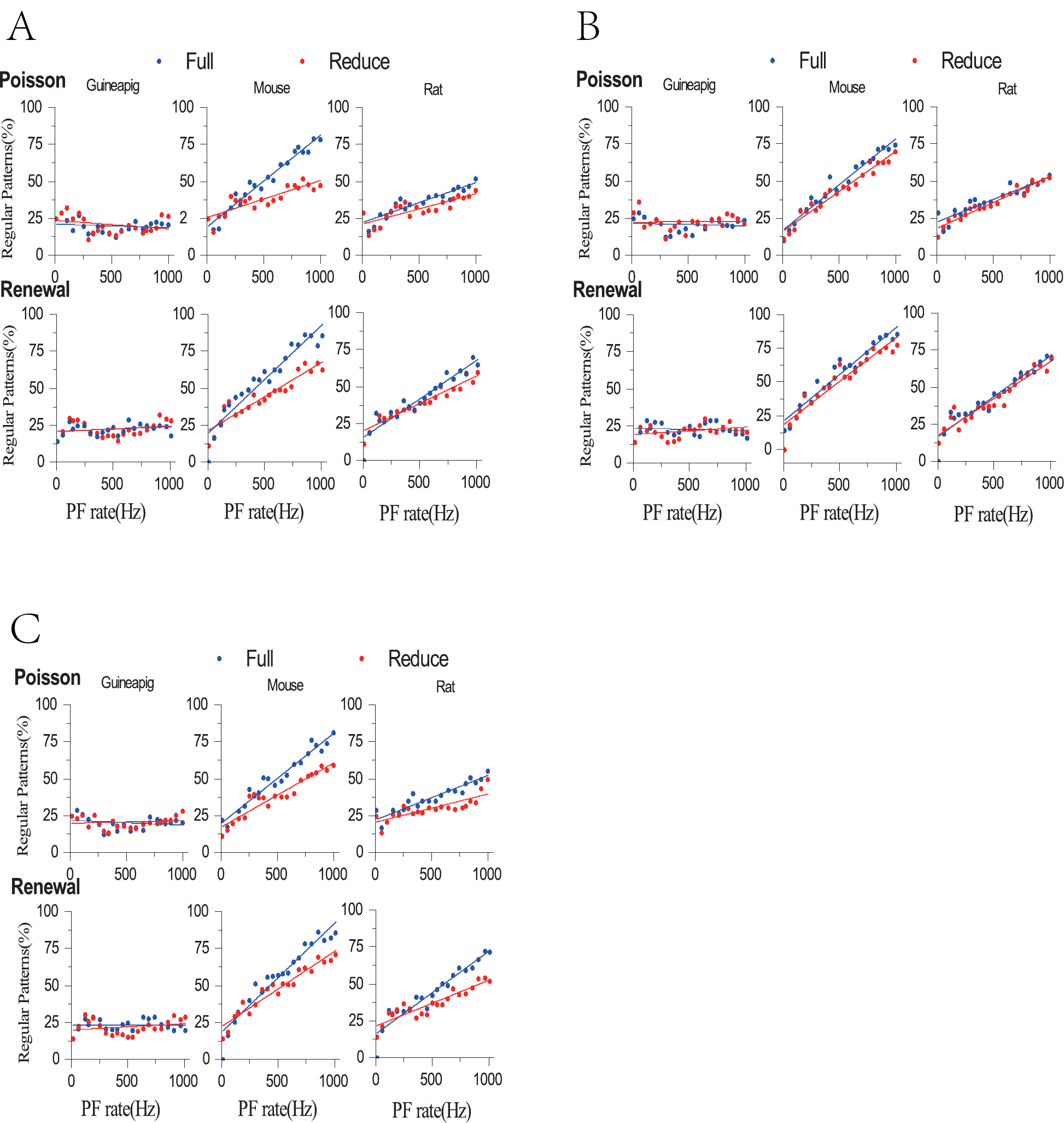}
    \label{}
\caption{Percentage of regular patterns in full (blue) and reduced model (red) with Poisson and renewal stimulation at different stimulation frequencies 10Hz to 1KHz. (A), (B) and (C) show Horton, Elect and Shreve reduced model respectively.}
\end{figure*}

\begin{figure*}[thbp]
 \centering	
\includegraphics[width=\columnwidth]{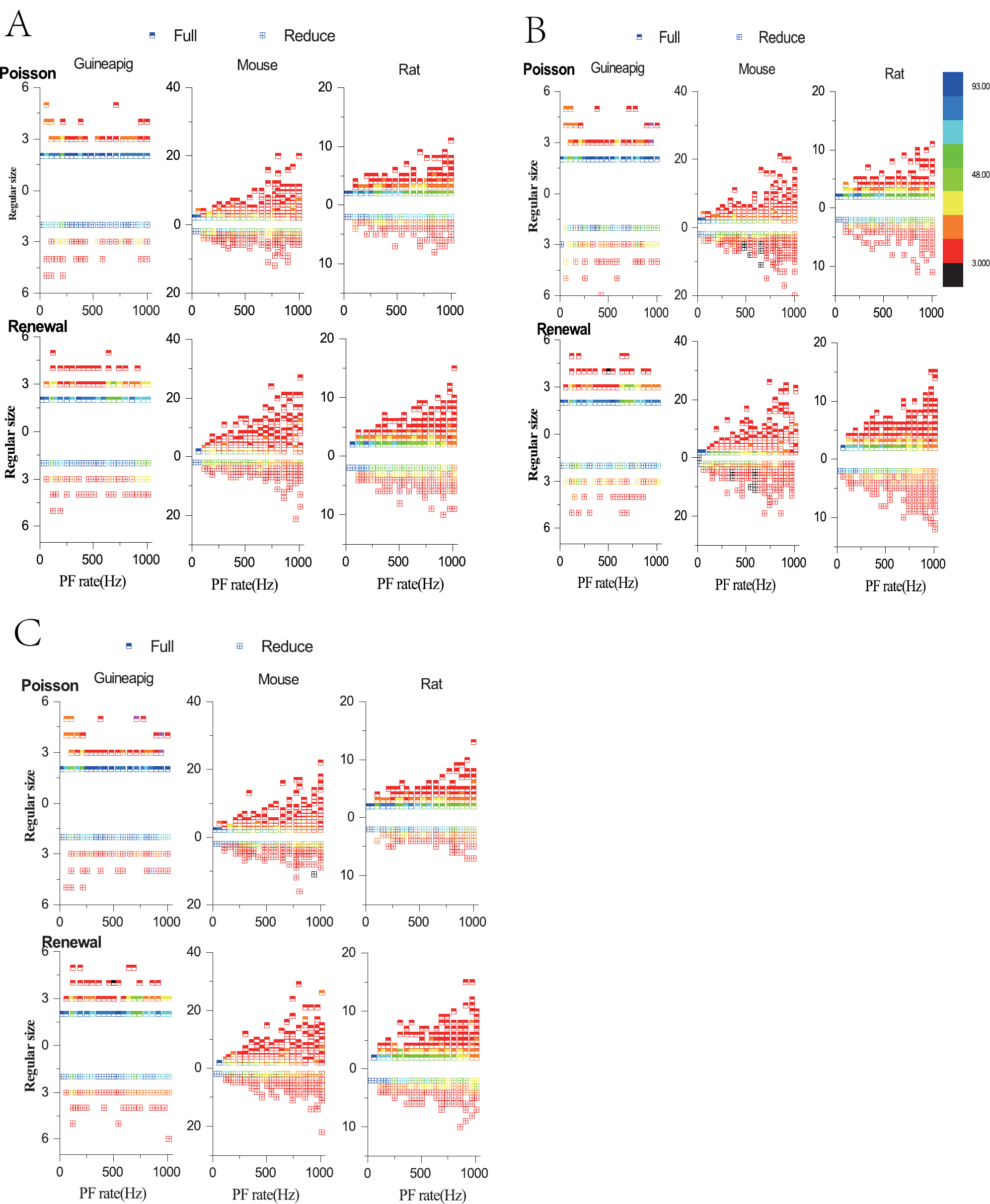}
    \label{}
\caption{Statistics of regular pattern size across a range of Poisson and renewal process stimulation for guinea-pig, mouse, and rat in full and reduced model. Percentage of different size is indicated by different colors, as there are more patterns in higher frequency. (A), (B) and (C) show Horton, Elect and Shreve reduced model respectively.}
\end{figure*}

\begin{figure*}[thbp]
 \centering	
\includegraphics[width=\columnwidth]{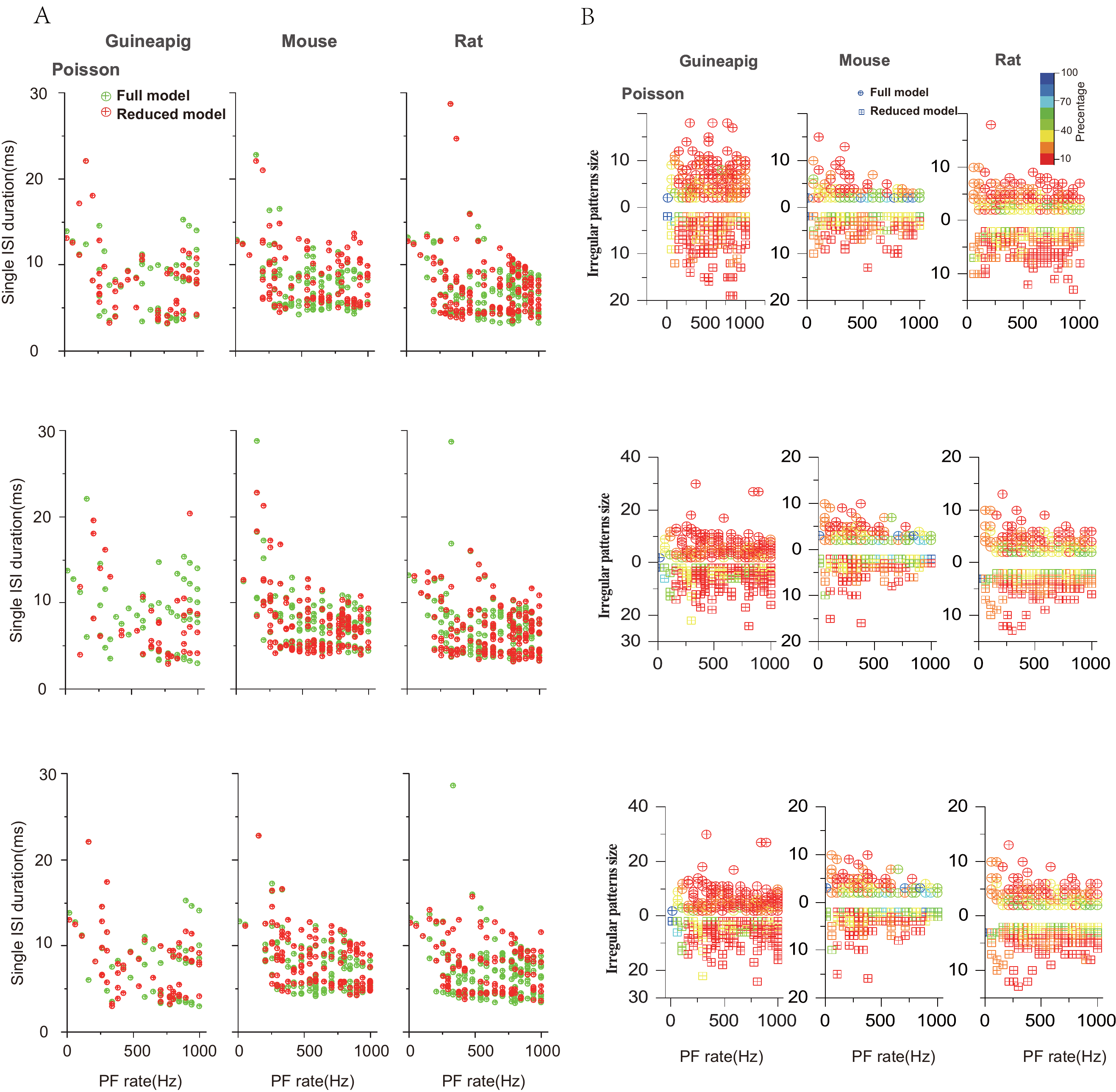}
    \label{}
\caption{Statistics of single ISI duration distribution and irregular pattern size across a range of Poisson  stimulation for guinea-pig, mouse, and rat in both full and reduced model. Percentage of different size is indicated by different colors.  Single size (A),  irregular size (B) of Horton (top), Elect (middle) and Shreve (bottom) reduced model, respectively.}
\end{figure*}


\begin{figure*}[thbp]
 \centering	
\includegraphics[width=\columnwidth]{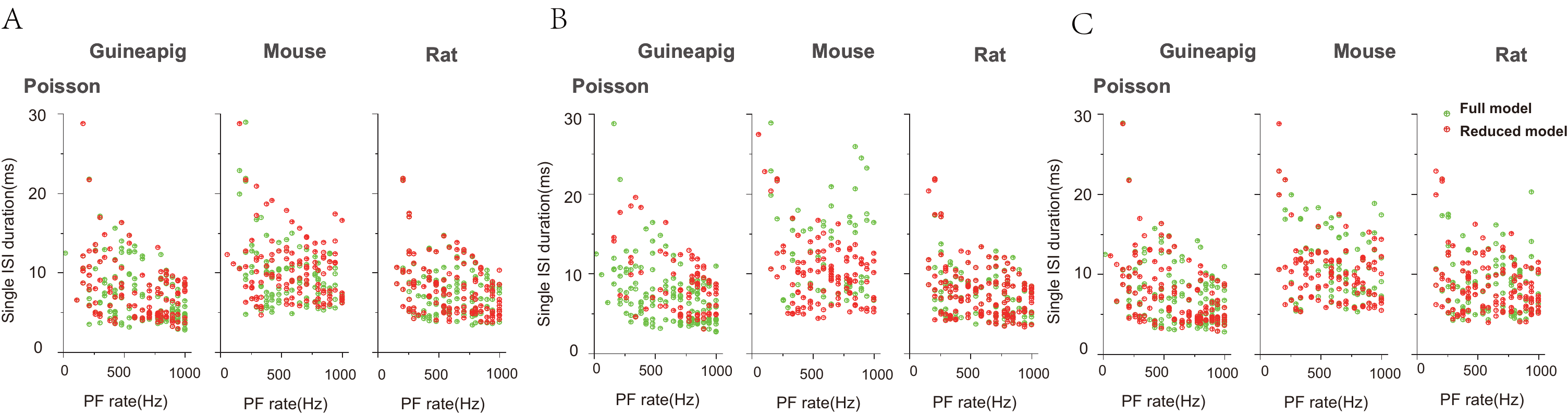}
    \label{}
\caption{Statistics of single ISI duration distribution across a range of Poisson  stimulation for guinea-pig, mouse, and rat in full and reduced model with inhibition input. (A) Single ISI duration distribution of Horton reduced model.  (B) Single ISI duration distribution of Elect reduced model. (C) Single ISI duration distribution of Shreve reduced model. }
\end{figure*}

\begin{figure*}[thbp]
 \centering	
\includegraphics[width=\columnwidth]{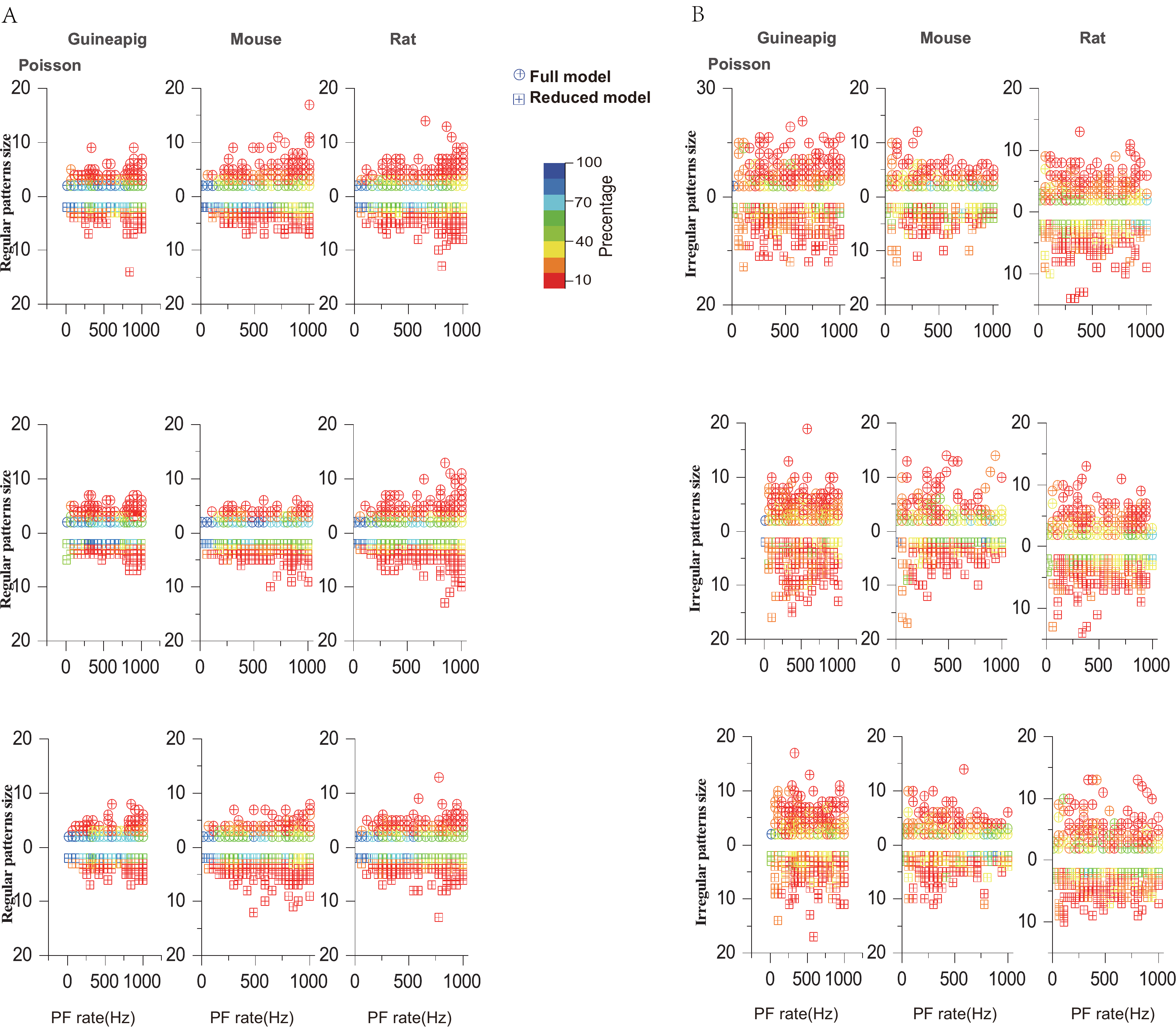}
    \label{}
\caption{Statistics of regular pattern size and irregular pattern across a range of Poisson  stimulation for guinea-pig, mouse, and rat in full and reduced model with inhibition input. Percentage of different size is indicated by different colors.  (A) Regular pattern size of Horton (top), Elect (middle) and Shreve (bottom) reduced model. (B) Irregular pattern size of Horton (top), Elect (middle) and Shreve (bottom) reduced model.}
\end{figure*}

\begin{figure*}[thbp]
\includegraphics[width=\columnwidth]{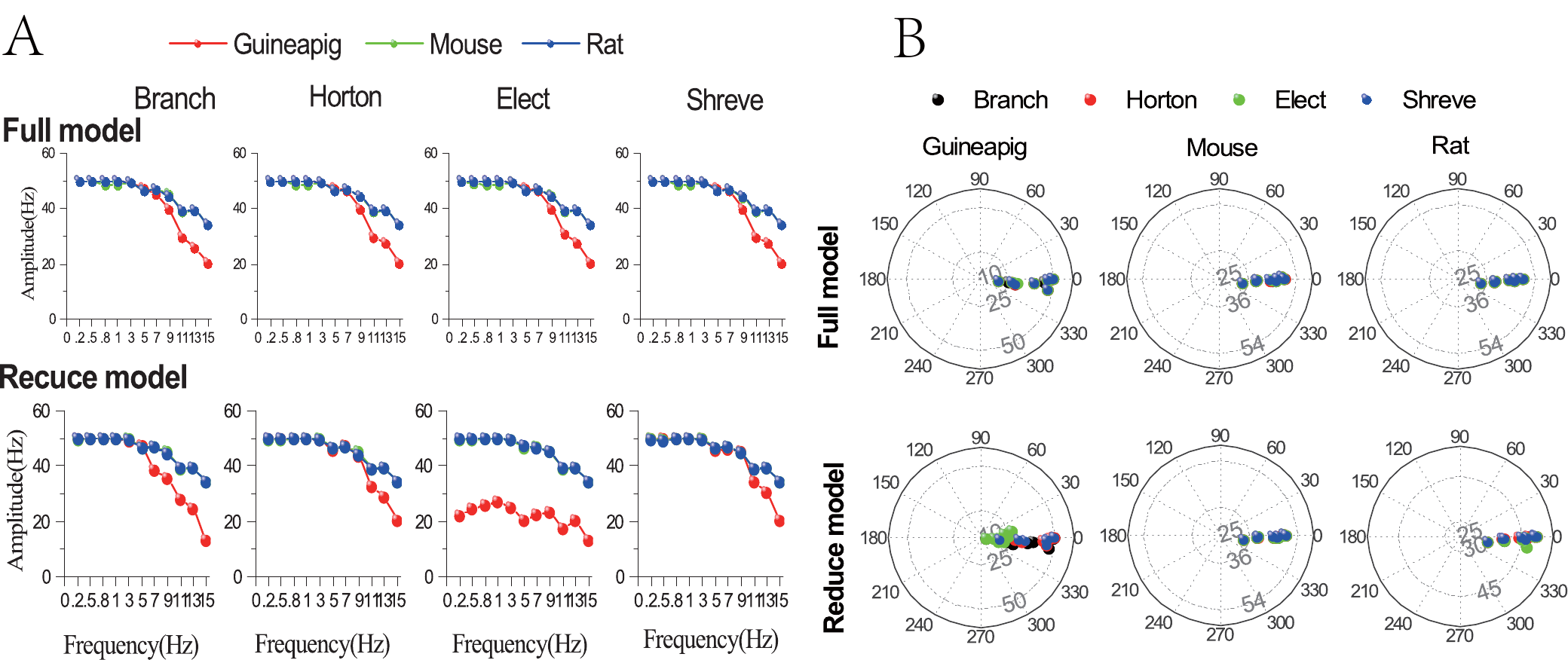}
    \centering	
    \label{}
\caption{ (A) Modulation amplitudes of PC output in full and reduced model of four reduced schemes for guinea-pig (red), mouse (green), and rat (blue), respectively, at different PF input frequencies.
    (B) Similar to (A), but for phase change of PC firing modulation for full and reduced model of four reduce schemes.  PF sinusoidal stimulation amplitude is 50Hz. }
\end{figure*}

\bibliographystyle{IEEEtran}
